\documentclass[acmsmall]{acmart}
\setcopyright{none}
\usepackage{booktabs}   %% For formal tables:
\usepackage{subcaption} %% For complex figures with subfigures/subcaptions
\begin{document}

%% Title information
\title[Dep-Fold]{Dependently Typed Folds for Nested Data Types}         %% [Short Title] is optional;
                                        %% when present, will be used in
                                        %% header instead of Full Title.
%% \titlenote{with title note}             %% \titlenote is optional;
                                        %% can be repeated if necessary;
                                        %% contents suppressed with 'anonymous'
%% \subtitle{Subtitle}                     %% \subtitle is optional
%% \subtitlenote{with subtitle note}       %% \subtitlenote is optional;
                                        %% can be repeated if necessary;
                                        %% contents suppressed with 'anonymous'

%% Author information
%% Contents and number of authors suppressed with 'anonymous'.
%% Each author should be introduced by \author, followed by
%% \authornote (optional), \orcid (optional), \affiliation, and
%% \email.
%% An author may have multiple affiliations and/or emails; repeat the
%% appropriate command.
%% Many elements are not rendered, but should be provided for metadata
%% extraction tools.

%% Author with single affiliation.
\author{Peng Fu}
%\authornote{with author1 note}          %% \authornote is optional;
                                        %% can be repeated if necessary
%\orcid{nnnn-nnnn-nnnn-nnnn}             %% \orcid is optional
\affiliation{
  %% \position{Position1}
  \department{Department of Mathematics and Statistics}              %% \department is recommended
  \institution{Dalhousie University}            %% \institution is required
  %% \streetaddress{Street1 Address1}
  %% \city{City1}
  %% \state{State1}
  %% \postcode{Post-Code1}
  %% \country{Country1}                    %% \country is recommended
}
\email{peng.frank.fu@gmail.com}          %% \email is recommended

%% Author with two affiliations and emails.
\author{Peter Selinger}
%\authornote{with author2 note}          %% \authornote is optional;
                                        %% can be repeated if necessary
%\orcid{nnnn-nnnn-nnnn-nnnn}             %% \orcid is optional
\affiliation{
  %% \position{Position2a}
  \department{Department of Mathematics and Statistics}             %% \department is recommended
  \institution{Dalhousie University}           %% \institution is required
  %% \streetaddress{Street2a Address2a}
  %% \city{City2a}
  %% \state{State2a}
  %% \postcode{Post-Code2a}
  %% \country{Country2a}                   %% \country is recommended
}
\email{selinger@mathstat.dal.ca}         %% \email is recommended
%% \affiliation{
%%   \position{Position2b}
%%   \department{Department2b}             %% \department is recommended
%%   \institution{Institution2b}           %% \institution is required
%%   \streetaddress{Street3b Address2b}
%%   \city{City2b}
%%   \state{State2b}
%%   \postcode{Post-Code2b}
%%   \country{Country2b}                   %% \country is recommended
%% }
%% \email{first2.last2@inst2b.org}         %% \email is recommended

%% Abstract
%% Note: \begin{abstract}...\end{abstract} environment must come
%% before \maketitle command
\begin{abstract}
  We present an approach to develop folds for nested data types using
  dependent types. We call such folds \textit{dependently typed folds},
  they have the following properties.
 (1) Dependently typed folds are defined by well-founded recursion and they
  can be defined in a total dependently typed language.
 (2) Dependently typed folds do not depend on maps, map functions and many terminating functions can be defined using dependently typed folds.
  (3) The induction principles
  for nested data types follow from the definitions of dependently typed folds and the programs defined by dependently typed folds can be formally verified.
  (4) Dependently typed folds exist for any nested data types and they can be specialized to the traditional \textit{higher-order folds}.
  Using various of examples, we show how to program and reason
  about dependently typed folds. We also show how to obtain dependently typed
  folds 
  in general and how to specialize them to the corresponding
  higher-order folds.

\end{abstract}

%% 2012 ACM Computing Classification System (CSS) concepts
%% Generate at 'http://dl.acm.org/ccs/ccs.cfm'.

%% End of generated code

%% Keywords
%% comma separated list
  %% \keywords are mandatory in final camera-ready submission

%% \maketitle
%% Note: \maketitle command must come after title commands, author
%% commands, abstract environment, Computing Classification System
%% environment and commands, and keywords command.
\maketitle

\section{Introduction}
\label{intro}
Consider the following list data type and its fold function in Agda\footnote{Agda documentation: \url{https://agda.readthedocs.io/en/v2.5.3/}.}.

\begin{verbatim}
data List (a : Set) : Set where
  Nil : List a
  Cons : a -> List a -> List a

fold : {a p : Set} -> p -> (a -> p -> p) -> List a -> p
fold {a} {p} base step Nil = base
fold {a} {p} base step (Cons x xs) = step x (fold {a} {p} base step xs)
\end{verbatim}

\noindent The key word \texttt{Set} is a kind that classifies types.
The function \texttt{fold} has two implicit type arguments, they correspond to the implicitly quantified type variables \texttt{a} and \texttt{p} in the type of \texttt{fold}. In Agda, implicit arguments can be supplied explicitly using braces (e.g. \texttt{\string{a\string}}), sometimes they can be omitted. 

The function \texttt{fold} is defined by structural recursion and it is terminating.  
Once \texttt{fold} is
defined, we can use it to define terminating
functions such as the following \texttt{map} and \texttt{sum}. This is similar to
 using the iterator to define terminating arithmetic functions in System \textbf{T} \cite[\S 7]{girard1989proofs}.

\begin{verbatim}
map : {a b : Set} -> (a -> b) -> List a -> List b
map f l = fold Nil (\ a r -> Cons (f a) r) l
sum : List Nat -> Nat
sum l = fold Z (\ x r -> add x r) l
\end{verbatim}

When defining the \texttt{map} function, if the input list \texttt{l} is empty, then we just return \texttt{Nil}, so the first argument for \texttt{fold} is \texttt{Nil}. If the input list \texttt{l} is of the form \texttt{Cons a as}, then we want to return \texttt{Cons (f a) (map f as)}, so the second argument for \texttt{fold} is \texttt{\string\ a r -> Cons (f a) r}, where \texttt{r} represents the result of the recursive call \texttt{map f as}. The function \texttt{sum} is defined similarly.

We can generalize the type of \texttt{fold} to obtain the following induction principle for list.

\begin{verbatim}
ind : {a : Set} {p : List a -> Set} -> p Nil ->
          ((x : a) -> {xs : List a} -> p xs -> p (Cons x xs)) -> (l : List a) -> p l
ind {a} {p} base step Nil = base
ind {a} {p} base step (Cons x xs) = step x {xs} (ind {a} {p} base step xs)
\end{verbatim}
Ignoring the implicit arguments, the definition of \texttt{ind} is the same as \texttt{fold}. Compared to the type of \texttt{fold}, the type of \texttt{ind} is more general as the kind of \texttt{p} is generalized from \texttt{Set} to \texttt{List a -> Set}, we call such \texttt{p} a \textit{property} of the list. The second argument \texttt{step} for \texttt{ind}
 has an implicit term argument \texttt{\string{xs : List a\string}}.
The induction principle \texttt{ind} states that to prove a property \texttt{p} holds for any
list \texttt{l}, one has to first prove that \texttt{Nil} has the property \texttt{p},
and then assuming \texttt{p} holds for any list \texttt{xs} as the induction hypothesis, prove that \texttt{p} holds
for \texttt{Cons x xs} for any \texttt{x}. 

%% If we erased all the implicit argument and the dependent indices of the type of \texttt{ind},
%% we get the type for \texttt{fold}.
We can now use the induction principle \texttt{ind} to prove that \texttt{map} behaves the same as the usual recursively defined \texttt{map'} (\texttt{lemma1}).

\begin{verbatim}
map' : {a b : Set} -> (a -> b) -> List a -> List b
map' f Nil = Nil
map' f (Cons x xs) = Cons (f x) (map' f xs)

lemma1 : {a b : Set} -> (f : a -> b) -> (l : List a) -> map f l == map' f l 
lemma1 {a} {b} f l =  
   ind {a} {\ y -> map f y == map' f y} refl (\ x {xs} ih -> cong (Cons (f x)) ih) l

refl : {a : Set} -> {x : a} -> x == x
cong : {a b : Set} -> {m n : a} -> (f : a -> b) -> m == n -> f m == f n
\end{verbatim}

In the proof of \texttt{lemma1}, we use a notion of equality \texttt{==} with 
the tactic \texttt{refl} to construct a reflexivity proof and the tactic
\texttt{cong} to construct
a congruence proof. The key to use the induction principle \texttt{ind} is to specify what kind of property on list we want to prove. In this case the property we have in mind is \texttt{\string\ y -> map f y == map' f y}.  Thus the type of \texttt{ind \string{a\string} \string{\string\ y -> map f y == map' f y \string}} is the following.

\begin{verbatim}
  map f Nil == Nil ->
  ((x : a) -> {xs : List a} -> map f xs == map' f xs ->
        map f (Cons x xs) == Cons (f x) (map' f xs)) ->
  (l : List a) -> map f l == map' f l
\end{verbatim}
The first two arguments for the above type correspond to the base case and the step case in the inductive proof. For the base case, we just need \texttt{refl}. For the step case, the induction
hypothesis \texttt{ih} has the type \texttt{map f xs == map' f xs}, we just need to
show \texttt{map f (Cons x xs) == Cons (f x) (map' f xs)}. Since \texttt{map f (Cons x xs)} can be evaluated to \texttt{Cons (f x) (map f xs)}, we finish the proof by a congruence on the induction hypothesis, so \texttt{cong (Cons (f x)) ih} is the
proof for \texttt{map f (Cons x xs) == Cons (f x) (map' f xs)}.

To summarize, the fold functions for \textit{regular} data types (i.e. non-nested inductive data types such as \texttt{List} and \texttt{Nat}) are well-behaved in the following sense. (1)
The fold functions are defined by well-founded recursion. (2) The fold functions can be used to define a range of terminating functions (including maps). (3) The types of the fold functions
can be generalized to the corresponding induction principles.
Unfortunately, these properties of folds for regular data types does not directly carry over to nested data types. Consider the following nested data type.

\begin{verbatim}
data Bush (a : Set) : Set where
  NilB : Bush a
  ConsB :  a -> Bush (Bush a) -> Bush a
\end{verbatim}
According to Bird and Meertens \cite[\S 1]{bird1998nested}, at each step down the list, entries are \textit{bushed}. For example, a value of type \texttt{Bush Nat} can be visualized as the following.
{
\begin{verbatim}

 bush1 = [ 4,                                        -- Nat
           [ 8, [ 5 ], [ [ 3 ] ] ],             -- Bush Nat
           [ [ 7 ], [ ], [ [ [ 7 ] ] ] ], -- Bush (Bush Nat)
           [ [ [ ], [ [ 0 ] ] ] ]  -- Bush (Bush (Bush Nat))
         ]           -- Bush (Bush (Bush (Bush (Bush Nat))))
\end{verbatim}
}
We can use general recursion to
define the following map function and fold function.%% \footnote{Here in Agda we need to turn on the unsafe \texttt{--no-termination} flag.}

\begin{verbatim}
hmapB : {b c : Set} -> (b -> c) -> Bush b -> Bush c
hmapB f NilB = NilB
hmapB f (ConsB x xs) = ConsB (f x) (hmapB (hmapB f) xs)

hfoldB : {a : Set} -> {p : Set -> Set} -> 
          ({b : Set} -> p b) -> ({b : Set} -> b -> p (p b) -> p b) -> Bush a -> p a
hfoldB base step NilB = base
hfoldB base step (ConsB x xs) = 
  step x (hfoldB base step (hmapB (hfoldB base step) xs))
\end{verbatim}

The fold function \texttt{hfoldB} for the nested data type \texttt{Bush} is
called a \textit{higher-order fold} in the literature (e.g. \cite{bird1999generalised}, \cite{johann2007initial}). 
Observe that the type variable \texttt{p} in \texttt{hfoldB} is generalized to kind \texttt{Set -> Set}. 
%% Although higher-order folds exists for any nested data types \cite[\S 3.2]{bird1999generalised},  there are at least the following four problems when using them. 

The higher-order fold \texttt{hfoldB} has the following problems. (1) The definition of \texttt{hfoldB} requires the map function \texttt{hmapB}, and \texttt{hmapB} can not be defined from \texttt{hfoldB}. (2) For both \texttt{hmapB} and \texttt{hfoldB}, Agda's termination checker fails because the use of recursion in both cases are not well-founded. For example, in the second
case of \texttt{hmapB}, the outer recursive call of \texttt{hmapB} is on the structurally smaller
argument \texttt{xs}, but there is no such indication for the inner recursive
call \texttt{(hmapB f)}. (3) Although possible (see Section \ref{related}), it is not immediately clear how \texttt{hfoldB} can be used to define functions such as a summation of all the entries in \texttt{Bush Nat}. The most obvious way is to instantiate \texttt{p} with \texttt{\string\ x -> Nat} for \texttt{hfoldB}, thus \texttt{hfoldB \string{Nat\string} \string{\string\ x -> Nat\string}} has type \texttt{(\string{b : Set\string} -> Nat) -> (\string{b : Set\string} -> b -> Nat -> Nat) -> Bush Nat -> Nat}.
We will have to provide a function of the type \texttt{(\string{b : Set\string} -> b -> Nat -> Nat)} as the second argument for \texttt{hfoldB \string{Nat\string} \string{\string\ x -> Nat\string}}. Such function would need to be defined for any type \texttt{b}, so it cannot be useful for defining the summation. (4) Unlike the induction principle for list, it is not clear how to
obtain an induction principle for \texttt{Bush} from the higher-order fold \texttt{hfoldB}. 

%% that what an induction principle is for \texttt{Bush} and what connection \texttt{hfoldB} has to the induction principle. 

As a result, in the pioneer works of \textit{generalized folds} for nested data types by Bird and Paterson (\cite{bird1999generalised}, \cite{bird1999bruijn}), they have to define
generalized folds using maps, and both generalized folds and map functions are defined by general recursion. %% Moreover, they assume working in a terminating fragment of Haskell, but it is unclear to
%% us how to characterize such fragment as it involves the use of general recursion. 
As for the formal verification, nested data types such as \texttt{Bush} can not be declared directly in the dependently typed language Coq\footnote{Coq user manual: \url{https://coq.inria.fr/refman/}}. In Agda, although we can declare nested data types such as \texttt{Bush}, we cannot easily program and reason about higher-order folds. 
This is because: (1) The Agda termination checker fails to recognize the termination of higher-order folds defined from general recursion. (2) Currently there is no natural formulation of induction principles for nested data types similar to the ones for regular data types.
 %% Moreover, it is unclear
%% how to reason about programs involving nested data types in Agda. 

%% and the termination of higher-order folds is not obvious, programming and reasoning about nested data types in total dependently typed languages such as Agda, Coq are quite challenging.
%a nontrivial problem (\cite{abel2003generalized}, \cite{matthes2009induction}). 
\subsection{Contributions of the paper}

We present an approach to define fold functions for nested data types using
dependent types inside the total dependently typed language Agda. We call such folds \textit{dependently typed folds}. Dependently typed folds are defined by well-founded recursion, hence
their termination is confirmed by Agda. Map functions and many other terminating functions can be defined directly from the dependently typed folds. Moreover, the higher-order folds (such
as \texttt{hfoldB}) are definable from the dependently typed folds. From the definitions of dependently typed folds, we can also obtain the corresponding induction principles. Thus we can formally reason about the programs involving nested data types in a total dependently typed language.
In this paper, we illustrate these ideas by focusing on several concrete examples, we 
also show how to obtain dependently typed folds in general.
%% and how to specialize them to the corresponding higher-order folds.

The main technical contents of the paper
are the following: In Section~\ref{dev}, using the \texttt{Bush} data type, we develop the first example of dependently typed folds. In Section~\ref{term} and Section~\ref{terme}, 
as a case study, we show how to define, program and reason about dependently typed folds
 in Agda using two well-known nested data types from the literature. In Section \ref{arbitrary}, we give an example to show how to obtain dependently typed folds in
general, and we show how to specialize dependently typed folds to the higher-order folds. In Section \ref{related}, we discuss related work. In Section \ref{final}, we discuss future work and conclude the paper.
All the detailed programs for each section are available at \url{https://github.com/Fermat/dependent-fold} and checked by Agda 2.5.3.

\section{A development of dependently typed folds in a total type theory via the Bush data type}
\label{dev}

%% In this section, we show how to program and reason about dependently typed folds using
%% three well-known examples from the literature. They are the \texttt{Bush} data type we
%% mentioned in the introduction (Section \ref{bush}), the data type \texttt{Term} for representing
%% the de Brujin index lambda terms (Section \ref{term}) and \texttt{TermE} for the improved version of \texttt{Term} (Section \ref{terme}). In Section \ref{index}, we show how dependently typed folds
%% give rise to the indexed data types called \textit{indexed representations} and how to obtain
%% the Church encodings of the indexed representations.
Let us continue the consideration of the \texttt{Bush} data type.  
The following is the result of evaluating \texttt{hmapB f bush1}, where \texttt{f : Nat -> b} for
some type \texttt{b}.

{
\begin{verbatim}
  [ f 4,                                            -- b
    [ f 8, [ f 5 ], [ [ f 3 ] ] ],             -- Bush b
    [ [ f 7 ], [ ], [ [ [ f 7 ] ] ] ],  -- Bush (Bush b)
    [ [ [ ], [ [ f 0 ] ] ] ]     -- Bush (Bush (Bush b))
  ]                -- Bush (Bush (Bush (Bush (Bush b))))
\end{verbatim}
}
\noindent In order to define the map function for \texttt{Bush Nat}, we need to already have the map functions defined for $\texttt{Bush}^n \ \texttt{Nat}$ for all $n \geq 0$, which seems paradoxical.
A way out is to define a general map function for $\texttt{Bush}^n$, for all $n \geq 0$. First we define $\texttt{Bush}^n$ as the following \texttt{NBush}. 

\begin{verbatim}
NTimes : (Set -> Set) -> Nat -> Set -> Set
NTimes p Z s = s
NTimes p (S n) s = p (NTimes p n s)
NBush : Nat -> Set -> Set
NBush = NTimes Bush
\end{verbatim}

\noindent The function \texttt{NTimes} is a type level function defined by pattern-matching on the natural number \texttt{n}. The function call \texttt{NTimes p n a} returns a type of the form $\texttt{p}^n\ a$. We now define the following
map function \texttt{mapB} for $\texttt{Bush}^n$,  where \texttt{mapB (S Z)} corresponds to the map function for \texttt{Bush a}.

\begin{verbatim}
mapB : {a b : Set} -> (n : Nat) -> (a -> b) -> NBush n a -> NBush n b 
mapB Z f x = f x
mapB (S n) f NilB = NilB
mapB (S n) f (ConsB x xs) = ConsB (mapB n f x) (mapB (S (S n)) f xs) 
\end{verbatim}

 The recursive definition of \texttt{mapB} is well-founded as all the recursive calls are on the components of the constructor \texttt{ConsB}. The Agda termination checker accepts this definition of \texttt{mapB}. The definition of \texttt{mapB} confirms a general principle in theorem proving: proving a more general lemma may be easier than proving a concrete one. We will use this principle again and again  when verifying programs involving nested data types.

From now on, instead
of considering the concrete data type \texttt{Bush}, we will focus on its generalized counterpart \texttt{NBush n}. Looking at the definition of \texttt{mapB}, we can view \texttt{NBush n} as a kind of abstract indexed data type that has three constructors. The first constructor has type \texttt{a -> NBush Z a}, corresponding to the first case of \texttt{mapB}.
The second constructor \texttt{NilB} has type \texttt{NBush (S n) a}, and the third constructor \texttt{ConsB} has type \texttt{NBush n a -> NBush (S (S n)) a -> NBush (S n) a}.

We now give the following dependently typed fold for the abstract indexed data type \texttt{NBush n}.

\begin{verbatim}
foldB : {a : Set} -> {p : Nat -> Set} ->
        (a -> p Z) ->
        ((n : Nat) -> p (S n)) ->
        ((n : Nat) -> p n -> p (S (S n)) -> p (S n)) ->
        (n : Nat) -> NBush n a -> p n
foldB base nil cons Z x = base x
foldB base nil cons (S n) NilB = nil n
foldB base nil cons (S n) (ConsB x xs) = 
  cons n (foldB base nil cons n x) (foldB base nil cons (S (S n)) xs)
\end{verbatim}

The dependently typed fold \texttt{foldB} captures the most general form of computing/traversal on the abstract data type \texttt{NBush n a}. The definition of \texttt{foldB} is well-founded because all the recursive calls
of \texttt{foldB} are on the components of \texttt{ConsB}. Observe that the definition of \texttt{foldB} has the same structure as \texttt{mapB}. We can redefine \texttt{mapB} using \texttt{foldB}.

\begin{verbatim}
mapB : {a b : Set} -> (n : Nat) -> (a -> b) -> NBush n a -> NBush n b
mapB {a} {b} n f l = 
  foldB {a} {\ n -> NBush n b} f (\ n -> NilB) (\ n -> ConsB) n l
\end{verbatim}

The dependently typed fold \texttt{foldB} allows us to directly define other terminating functions such as the summation of all the entries in \texttt{Bush Nat} and the length function for \texttt{Bush}. 

\begin{verbatim}
sumB : Bush Nat -> Nat
sumB = foldB {Nat} {\ n -> Nat} (\ x -> x) (\ n -> Z) (\ n -> add) (S Z) 

lengthB : {a : Set} -> Bush a -> Nat
lengthB {a} = foldB {a} {\ n -> Nat} (\ x -> Z) (\ n -> Z) (\ n r1 r2 -> S r2) (S Z)
\end{verbatim}

Comparing the dependently typed fold \texttt{foldB} and the higher-order fold \texttt{hfoldB} in Section~\ref{intro}, we can see that \texttt{foldB} does not depend on map, and \texttt{mapB} can be defined from \texttt{foldB}. The termination of \texttt{foldB} is obvious and it can be used
 to define other terminating functions. Moreover, the higher-order fold \texttt{hfoldB} is an instance of the dependently typed fold \texttt{foldB}, as we can define \texttt{hfoldB} using \texttt{foldB}.

\begin{verbatim}
hfoldB : {a : Set} -> {p : Set -> Set} -> 
          ({b : Set} -> p b) -> ({b : Set} -> b -> p (p b) -> p b) -> Bush a -> p a
hfoldB {a} {p} base step = 
  foldB {a} {\ n -> NTimes p n a} (\ x -> x) (\ n -> base) (\ n -> step) (S Z)
\end{verbatim}
Last but not least, we can generalize the type of dependently typed fold \texttt{foldB} to obtain the following induction principle \texttt{indB}, just like how we obtain the induction principle for \texttt{List} from its fold function.

\begin{verbatim}
indB : {a : Set} -> {p : (n : Nat) -> NBush n a -> Set} -> 
       ((x : a) -> p Z x) -> 
       ((n : Nat) -> p (S n) NilB) ->
       ((n : Nat) -> {x : NBush n a} -> {xs : NBush (S (S n)) a} ->
               p n x -> p (S (S n)) xs -> p (S n) (ConsB x xs)) ->
       (n : Nat) -> (xs : NBush n a) -> p n xs
indB base nil cons Z xs = base xs
indB base nil cons (S n) NilB = nil n
indB base nil cons (S n) (ConsB x xs) =
  cons n (indB base nil cons n x) (indB base nil cons (S (S n)) xs)
\end{verbatim}

Observe that the definition of \texttt{indB} is the same as \texttt{foldB}, and
the type variable \texttt{p} is generalized to kind \texttt{(n : Nat) -> NBush n a -> Set}. The
type of \texttt{indB} is specifying how to prove a property \texttt{p} holds for any \texttt{xs} of type \texttt{NBush n a} by induction.
More specifically, for the first base case, we need to show \texttt{p} holds for any \texttt{x} of type \texttt{NBush Z a} (which equals \texttt{a}), hence \texttt{p Z x}. For the second base case, we need to show \texttt{p} holds for \texttt{NilB} of type \texttt{NBush (S n) a}. For the step case, we assume \texttt{p} holds for \texttt{x} of type \texttt{NBush n a} and \texttt{xs} of type \texttt{NBush (S (S n)) a} as the inducton hypotheses, we need to show
\texttt{p} holds for \texttt{ConsB x xs}. 

With \texttt{indB}, we can now prove properties about \texttt{mapB} and \texttt{foldB}. In the following we prove that \texttt{mapB} has the usual identity and composition properties. 

\begin{verbatim}
identity : {a : Set} -> (n : Nat) -> (y : NBush n a) -> y == mapB n (\ x -> x) y
identity {a} n y =
  indB {a} {\ n v -> v == mapB n (\ x -> x) v} (\ x -> refl)
   (\ n -> refl) (\ n {x} {xs} ih1 ih2 -> cong2 ConsB ih1 ih2) n y

mapCompose : {a b c : Set} -> (n : Nat) -> (f : b -> c) -> (g : a -> b) ->
 (x : NBush n a) -> mapB n (compose f g) x == mapB n f (mapB n g x)
mapCompose {a} {b} {c} n f g x =
  indB {a} {\ n v -> mapB n (compose f g) v == mapB n f (mapB n g v)}
   (\ v -> refl) (\ n -> refl) (\ n {x1} {xs} ih1 ih2 -> cong2 ConsB ih1 ih2) n x

cong2 : {a b c : Set} -> {m1 n1 : a} -> {m2 n2 : b} -> 
         (f : a -> b -> c) -> m1 == n1 -> m2 == n2  -> f m1 m2 == f n1 n2
\end{verbatim}
%% compose : {a b c : Set} -> (b -> c) -> (a -> b) -> (a -> c)
%% compose g f = \ x -> g (f x)

We note that the usual way of proving things in Agda is by recursion, and relying on the Agda
termination checker to prove termination. However, since our purpose is to show the strength
of induction principles such as \texttt{indB}, we do not use recursion at all. All the proofs in this paper are by induction principles.

Let us take a closer look at \texttt{identity}. It is a general statement of
\texttt{mapB n} (includes the special case \texttt{mapB (S Z)}). It is also about
a property of
\texttt{y : NBush n a}, i.e. \texttt{y} has the property of being equal to \texttt{mapB n (\string\ x -> x) y} for any \texttt{n}. So we instantiate the property \texttt{p} in \texttt{indB} with 
\texttt{\string\ n v -> v == mapB n (\string\ x -> x) v}. Thus the type of
\texttt{indB \string{a\string} \string{\string\ n v -> v == mapB n (\string\ x -> x) v\string}}
is the following.

\begin{verbatim}
((x : a) -> x == mapB Z (\ x -> x) x) ->
((n : Nat) -> NilB == mapB (S n) (\ x -> x) NilB) ->
((n : Nat) -> {x : NBush n a} -> {xs : NBush (S (S n)) a} ->
      x == mapB n (\ x -> x) x ->
      xs == mapB (S (S n)) (\ x -> x) xs ->
      ConsB x xs == mapB (S n) (\ x -> x) (ConsB x xs)) ->
(n : Nat) -> (xs : NBush n a) -> xs == mapB n (\ x -> x) xs  
\end{verbatim}

\noindent We need to provide three arguments for \texttt{indB \string{a\string} \string{\string\ n v -> v == mapB n (\string\ x -> x) v\string}} to obtain a proof of the theorem \texttt{(n : Nat) -> (xs : NBush n a) -> xs == mapB n (\string\ x -> x) xs}. These three arguments correspond to the three cases in the inductive proof, i.e. a case for \texttt{NBush Z a},
a case for \texttt{NilB} and a case for \texttt{ConsB x xs}. In
the third case for \texttt{ConsB x xs}, we have two induction hypotheses, i.e. \texttt{ih1 : x == mapB n (\string\ x -> x) x} and \texttt{ih2 : xs == mapB (S (S n)) (\string\ x -> x) xs}. A congruence on these two induction hypotheses, i.e. \texttt{cong2 ConsB ih1 ih2},  gives us the proof for \texttt{ConsB x xs == mapB (S n) (\string\ x -> x) (ConsB x xs)}.
The proof of \texttt{mapCompose} is similar to the proof of \texttt{identity}.

Recall that Agda does not accept the general recursive definition of \texttt{hmapB} in Section \ref{intro}. Now that we understand
that \texttt{hmapB} is just \texttt{mapB (S Z)}, we can use the induction principle \texttt{indB}
to show that \texttt{mapB (S Z)} has the same computational behavior as \texttt{hmapB}.

\begin{verbatim}
mapNilB : forall {a b : Set} -> (f : a -> b) -> mapB (S Z) f NilB == NilB
mapNilB {a} {b} f = refl

mapConsB : {a b : Set} -> (f : a -> b) -> (x : a) -> (xs : Bush (Bush a)) ->
           mapB (S Z) f (ConsB x xs) == ConsB (f x) (mapB (S Z) (mapB (S Z) f) xs)
mapConsB {a} {b} f x xs = cong (ConsB (f x)) (addMap {a} {b} (S Z) f xs)

addMap : {a b : Set} -> (n : Nat) -> (f : a -> b) -> (x : NBush (add n n) a) ->
          mapB (add n n) f x == mapB n (mapB n f) x
addMap {a} {b} n f x  = 
  indB {NBush n a} {\ m v -> mapB (add m n) f v == mapB m (mapB n f) v}
       (\ x -> refl) (\ n -> refl) 
       (\ n {x} {xs} ih1 ih2 -> cong2 ConsB ih1 ih2) n x
\end{verbatim}
The theorem \texttt{mapNilB} corresponds to the first case in the general recursive definition of \texttt{hmapB},
the theorem \texttt{mapConsB} corresponds to the second case. To prove \texttt{mapConsB}, we need to prove a more general lemma \texttt{addMap}. The proof of lemma \texttt{addMap} is
by standard induction, however, coming up with the correct lemma \texttt{addMap} requires some effort. 

Similarly, we can 
use \texttt{indB} to show that the \texttt{hfoldB} defined from \texttt{foldB} behaves the same as the one defined by general recursion.
The following theorem \texttt{foldBNilB} corresponds to the first case in the general recursive definition, and theorem \texttt{foldBConsB} corresponds to the second case. We elide the nontrivial proof of the lemma \texttt{lemmConsB} (which uses \texttt{indB}).

\begin{verbatim}
foldBNilB : {a : Set} -> {p : Set -> Set} -> (base : {b : Set} -> p b) ->
            (step : {b : Set} -> b -> p (p b) -> p b) -> 
            hfoldB {a} {p} base step NilB == base
foldBNilB base step = refl

foldBConsB : {a : Set} -> {p : Set -> Set} -> (base : {b : Set} -> p b) ->
             (step : {b : Set} -> b -> p (p b) -> p b) ->
             (x : a) -> (xs : Bush (Bush a)) -> 
             hfoldB base step (ConsB x xs) ==
             step x (hfoldB base step (mapB (S Z) (hfoldB base step) xs))
foldBConsB {a} {p} base step x xs = 
  cong (step x) (lemmConsB {a} {p} (S Z) base step xs)
\end{verbatim}

Finally, we use the induction principle \texttt{indB} to prove that for any function \texttt{f}, if \texttt{f}
behaves according to \texttt{foldBNilB} and \texttt{foldBConsB}, i.e.
\texttt{f base step NilB == base} and \texttt{f base step (ConsB x xs) == step x (f base step (mapB (S Z) (f base step) xs)}, then \texttt{f} is equal to \texttt{hfoldB}. This statement
can be formalized as the following. 

{
\begin{verbatim}

uniqueness :  (f : {a : Set} -> {p : Set -> Set} -> 
                   ({b : Set} -> p b) -> 
                   ({b : Set} -> b -> p (p b) -> p b) -> Bush a -> p a) ->

              (hp1 : {a : Set} {p : Set -> Set} -> (base : {b : Set} -> p b) -> 
                     (step : {b : Set} -> b -> p (p b) -> p b) ->
                      f {a} {p} base step NilB == base) ->

              (hp2 : {a : Set} {p : Set -> Set} -> (base : {b : Set} -> p b) ->
                     (step : {b : Set} -> b -> p (p b) -> p b) ->
                     (x : a) -> (xs : Bush (Bush a)) ->
                     f base step (ConsB x xs) ==
                     step x (f base step (mapB (S Z) (f base step) xs))) ->

               {a : Set} -> {p : Set -> Set} ->        
               (base : {b : Set} -> p b) ->       
               (step : {b : Set} -> b -> p (p b) -> p b) -> (bush : Bush a) ->       
               f {a} {p} base step bush == hfoldB {a} {p} base step bush 
uniqueness f hp1 hp2 {a} {p} base step bush = 
  indB {a} {\ n v -> lift n (f base step) v == lift n (hfoldB base step) v} ...

lift : {a : Set} {p : Set -> Set}
         (n : Nat) -> (g : {a : Set} -> Bush a -> p a) -> NBush n a -> NTimes p n a
lift {a} {p} Z g x = x
lift {a} {p} (S n) g x = g (mapB (S Z) (lift {a} {p} n g) x)
\end{verbatim}
}
The proof of \texttt{uniqueness} is nontrivial and requires all the lemmas and
theorems about \texttt{foldB} and \texttt{mapB} we seen so far together with the helper function
\texttt{lift}. The complete proof can be found in the supplementary material. The key idea of the proof is that instead of proving the
concrete theorem \texttt{(bush : Bush a) -> f base step bush == hfoldB base step bush},
we use \texttt{lift} to prove a more general one, namely, \texttt{(n : Nat) -> (bush : NBush n a) -> lift n (f base step) bush == lift n (hfoldB base step) bush}. 

\subsection{The indexed representations}
\label{index}
We now show that the nested data type \texttt{Bush} is in fact definable even in a core type theory without nested data types. Indeed, we can define \texttt{NBush} directly as a non-nested
indexed data type \texttt{BushN} (called the \textit{indexed representation}). The nested data type \texttt{Bush} is recoverable as \texttt{BushN (S Z)}. This opens the possibility of a user defined nested data type in a surface language, then it can be automatically desugared to a
non-nested definition in the underlying type theory while still providing the fold function
and induction principle. For example, the total dependently typed language Coq does not accept nested data types such as \texttt{Bush} due to the failure of Coq's strict positivity check. So in this case we can work with the indexed representations instead.

Consider the following indexed data type \texttt{BushN}.

\begin{comment}
Recall the following definition of dependently typed fold \texttt{foldB}. 
{
\begin{verbatim}
foldB : {a : Set} -> {p : Nat -> Set} ->
        (a -> p Z) ->
        ((n : Nat) -> p (S n)) ->
        ((n : Nat) -> p n -> p (S (S n)) -> p (S n)) ->
        (n : Nat) -> NBush n a -> p n
foldB base nil cons Z x = base x
foldB base nil cons (S n) NilB = nil n
foldB base nil cons (S n) (ConsB x xs) = 
  cons n (foldB base nil cons n x) (foldB base nil cons (S (S n)) xs)
\end{verbatim}
}
\end{comment}

\noindent %Here \texttt{NBush} is a type-level function of kind \texttt{Nat -> Set -> Set}.
%% We mentioned that we can view \texttt{NBush n a}
%% as a kind of abstract indexed data type that has three constructors. In fact, we can internalize this
%% observation and make these three constructors explicit. 
%, its fold function and its induction principle.

\begin{verbatim}
data BushN : Nat -> Set -> Set where
  Base : {a : Set} -> a -> BushN Z a
  NilBN : {a : Set} -> {n : Nat} -> BushN (S n) a
  ConsBN : {a : Set} -> {n : Nat} -> 
            BushN n a -> BushN (S (S n)) a -> BushN (S n) a
\end{verbatim}
The \texttt{BushN} data type is indexed by the natural numbers.
A value of type \texttt{BushN Z a}
is of the form \texttt{Base x}, where \texttt{x : a}. A value of type \texttt{BushN (S n) a} can be either a \texttt{NilBN}, or \texttt{ConsBN x xs} with \texttt{x : BushN n a} and \texttt{xs : BushN (S (S n)) a}. %% Informally, the natural number is used to encode the \textit{nestedness} for \texttt{Bush}.

The following are the fold function and the induction principle for
\texttt{BushN}. 

\begin{verbatim}
foldBN : {a : Set} -> {p : Nat -> Set} ->
         (a -> p Z) ->
         ((n : Nat) -> p (S n)) ->
         ((n : Nat) -> p n -> p (S (S n)) -> p (S n)) ->
         (n : Nat) -> BushN n a -> p n
foldBN base nil cons Z (Base x) = base x
foldBN base nil cons (S n) NilBN = nil n
foldBN base nil cons (S n) (ConsBN x xs) = 
  cons n (foldBN base nil cons n x) (foldBN base nil cons (S (S n)) xs)

indBN : {a : Set} -> {p : (n : Nat) -> BushN n a -> Set} -> 
        ((x : a) -> p Z (Base x)) -> 
        ((n : Nat) -> p (S n) NilBN) ->
        ((n : Nat) -> {x : BushN n a} -> {xs : BushN (S (S n)) a} ->
              p n x -> p (S (S n)) xs -> p (S n) (ConsBN x xs)) ->
        (n : Nat) -> (xs : BushN n a) -> p n xs
\end{verbatim}
The definition of \texttt{indNB} is exactly the same as \texttt{foldBN}. 
Notice that \texttt{foldBN} is almost the same as \texttt{foldB} except it actually pattern-matches on all the constructors of the index data type \texttt{BushN n a}. We can convert back and forth
between \texttt{NBush n a} and \texttt{BushN n a}.

\begin{verbatim}
to : {a : Set} -> (n : Nat) -> NBush n a -> BushN n a
to {a} n s = 
  foldB {a} {\ n -> BushN n a} Base (\ n -> NilBN) (\ n -> ConsBN) n s
from : {a : Set} -> (n : Nat) -> BushN n a -> NBush n a
from {a} n s = 
  foldBN {a} {\ n -> NBush n a} (\ x -> x) (\ n -> NilB) (\ n -> ConsB) n s 

toFrom : {a : Set} -> (n : Nat) -> (x : NBush n a) -> from n (to n x) == x
fromTo : {a : Set} -> (n : Nat) -> (x : BushN n a) -> to n (from n x) == x
\end{verbatim}
In principle, all the programs and theorems about \texttt{NBush n a} can be converted to \texttt{BushN n a}. Please see the supplementary material for an example. 

%% We say \texttt{BushN} is \textit{the indexed representation} of the abstract indexed data type \texttt{NBush}. 

\subsection{The Church encodings of the indexed representations}
\label{church}

We can work with the indexed representations via their Church encodings in the Calculus of Constructions \cite{coquand1988calculus}, a minimal total dependent type system that does not provide primitive data types and recursion.  %We can obtain the Church encodings of the indexed representations.
As an example, we now define \texttt{CNBush}, the Church-encoded version of \texttt{BushN}. 

\begin{verbatim}
CNBush : Nat -> Set -> Set
CNBush n a = {p : Nat -> Set} ->
             (a -> p Z) ->
             ((n : Nat) -> p (S n)) ->
             ((n : Nat) -> p n -> p (S (S n)) -> p (S n)) -> p n
\end{verbatim}
The definition of \texttt{CNBush} is impredicative. The kind of \texttt{CNBush} should be $\texttt{Nat -> Set ->}\texttt{Set}_1$, not \texttt{Nat -> Set -> Set} (recall that \texttt{Set} in Agda is a shorthand for $\texttt{Set}_0$), due to the quantification of \texttt{p : Nat -> Set}.  Since Agda does not support impredicative polymorphism, we enable this feature by using the unsafe \texttt{--type-in-type} flag. In a language that supports impredicativity (e.g. Coq), defining the Church-encoded \texttt{CNBush} is not a problem, we also provide the Coq version of \texttt{CNBush} in the supplementary material.  

To obtain the Church-encoded \texttt{BushN}, we first identify the type \texttt{CNBush n a} with the type of \texttt{foldBN}, then we define the three constructors of \texttt{CNBush} by imitating the three cases of \texttt{foldBN}. 

\begin{verbatim}
cbase : {a : Set} -> a -> CNBush Z a
cbase x = \ base nil cons -> base x

cnil : {a : Set} -> (n : Nat) -> CNBush (S n) a
cnil n = \ base nil cons -> nil n

ccons : {a : Set} -> (n : Nat) -> CNBush n a -> CNBush (S (S n)) a -> CNBush (S n) a
ccons n x xs = \ base nil cons -> cons n (x base nil cons) (xs base nil cons)
\end{verbatim}

 Since the principle of fold is already encoded in the constructors, the following definition of \texttt{cfoldB} is essentially an identity function.

\begin{verbatim}
cfoldB : {a : Set} -> {p : Nat -> Set} ->
         (a -> p Z) ->
         ((n : Nat) -> p (S n)) ->
         ((n : Nat) -> p n -> p (S (S n)) -> p (S n)) ->
         (n : Nat) -> CNBush n a -> p n
cfoldB base nil cons n b = b base nil cons             
\end{verbatim}

Programming with the Church-encoded \texttt{CNBush} is just like
programming with \texttt{foldBN} for \texttt{BushN}. Instead of using \texttt{foldBN}, we
use \texttt{cfoldB}. For example, the following is the map function for \texttt{CNBush}.

\begin{verbatim}
cmapB : {a b : Set} -> (n : Nat) -> (a -> b) -> CNBush n a -> CNBush n b
cmapB {a} {b} n f = 
  cfoldB {a} {\ n -> CNBush n b} (\ x -> cbase (f x)) cnil ccons n  
\end{verbatim}
Note that we do not use any recursion in the definitions of \texttt{cfoldB} and \texttt{cmapB}.

Although we can program with the Church-encoded \texttt{CNBush} using \texttt{cfoldB},
we cannot obtain the induction principle from \texttt{cfoldB}, as it is well-known that
 induction is not derivable in the Calculus of Construction (\cite{coquand:inria-00075471}, \cite{geuvers2001}).  

 \section{Case study I: de Bruijn notation as the nested data type \texttt{Term}}
 \label{term}
One place in the literature where nested data types are useful is the representation of de Bruijn  lambda terms. In this section and the next section, we give an extended case study of two nested data types that are used to represent the de Bruijn lambda terms. 
 The case study demonstrates that the dependently typed folds is sufficient for the purpose of programming and reasoning about nested data types. 

%\subsection{De Bruijn notation as the nested data type \texttt{Term}}
%% We now show how to program and reason about dependently typed folds for other nested data types such as the ones defined by Bird and Paterson \cite{bird1999bruijn}.
Recall that the idea of de Bruijn notation is to use a number to represent a bound variable. The number is the number of binders between the bound variable and its binding site \cite{de1972lambda}. For example,
the lambda term $\lambda x . x \ (\lambda y . x \ y \ (\lambda z . x \ y \ z))$ is represented as $\lambda . 0 \ (\lambda . 1 \ 0 \ (\lambda . 2 \ 1 \ 0))$. This idea can be captured by
the following data types (from \cite{bird1999bruijn}). %% , where bound variables are represented by numbers, i.e. of
%% the type \texttt{Incr a} and free variables are represented by their names, i.e. of type \texttt{a}.

\begin{verbatim}
data Incr (a : Set) : Set where
  Zero : Incr a
  Succ : a -> Incr a

data Term (a : Set) : Set where
  Var : a -> Term a
  App : Term a -> Term a -> Term a
  Lam : Term (Incr a) -> Term a
\end{verbatim}

The data type \texttt{Term} is a nested data type because the constructor \texttt{Lam} requires
an argument of \textit{larger} type \texttt{Term (Incr a)}, instead of \texttt{Term a}. At each level down the constructor \texttt{Lam}, a term will gain an additional \texttt{Incr} in its type. For example, the following are the representations of the terms $\lambda . 0 \ (\lambda . 1 \ 0 \ (\lambda . 2 \ 1 \ 0))$ and $\lambda . \lambda . 1\ 0 \ (S (S `W`))$.

\begin{verbatim}
term1 : Term Char
term1 = Lam (App (Var Zero)                  -- Var Zero : Term (Incr Char)
                 (Lam (App (App (Var (Succ Zero)) 
                                (Var Zero))  -- Var Zero : Term (Incr (Incr Char))
                            (Lam (App (App (Var (Succ (Succ Zero)))
                                           (Var (Succ Zero)))
                                      (Var Zero)))))) 

term2 : Term Char
term2 = Lam (Lam (App (App (Var (Succ Zero)) (Var Zero))
                       (Var (Succ (Succ W)))))
                           -- Var (Succ (Succ W)) : Term (Incr (Incr Char))
\end{verbatim}

Notice that each variable in \texttt{term1}, \texttt{term2} has a type of the form $\texttt{Term}\ (\texttt{Incr}^n \ \texttt{Char})$. In a term of the type \texttt{Term Char}, the maximum number of \texttt{Succ} in a bound variable is strictly less than the number of \texttt{Incr} in its type, while the number of \texttt{Succ} in a free variable is equal to the number of \texttt{Incr} in its type. %% For another example, $\lambda . \underline{(S \ `W`)} \ (\lambda . 1 \ 0 \ (\lambda . 2 \ 1 \ 0))$ has type \texttt{Term Char}, but $\lambda . \underline{1} \ (\lambda . 1 \ 0 \ (\lambda . 2 \ 1 \ 0))$ has type \texttt{Term (Incr Char)}. 

%% Similar to the data type \texttt{Nested}, the following is
\subsection{Dependently typed folds for \texttt{Incr} and \texttt{Term}}
Since we will need to manipulate both bound and free variables, 
we define the following
dependently typed fold \texttt{foldI} and \texttt{mapIncr} function for $\texttt{Incr}^n\ a$. 

\begin{verbatim}
NIncr : Nat -> Set -> Set
NIncr = NTimes Incr

foldI : {a : Set} -> {p : Nat -> Set} -> (n : Nat) ->
        (a -> p Z) ->
        ((m : Nat) -> p (S m)) ->
        ((m : Nat) -> p m -> p (S m)) -> NIncr n a -> p n
foldI Z base zero succ x = base x
foldI (S n) base zero succ Zero = zero n
foldI (S n) base zero succ (Succ x) = succ n (foldI n base zero succ x)

mapIncr : {a b : Set} -> (n : Nat) -> (a -> b) -> NIncr n a -> NIncr n b
mapIncr {a} {b} n f y = 
  foldI {a} {\ n -> NIncr n b} n f (\ m -> Zero) (\ m -> Succ) y
\end{verbatim}
%%  Observe that \texttt{mapIncr} only map the
%% function \texttt{f} to the \texttt{x} where \texttt{x} is neither
%% \texttt{Succ} nor \texttt{Zero}.
Of course, the usual fold function for the regular data type \texttt{I} is an instance of \texttt{foldI}.

\begin{verbatim}
foldI' : {a : Set} -> {p : Set} -> p -> (a -> p) -> Incr a -> p
foldI' {a} {p} zero succ = foldI {a} {\ n -> p} (S Z) succ (\ m -> zero) (\ m x -> x)           
\end{verbatim}

Let \texttt{a} be a fixed constant type. We say \texttt{y : NIncr n a} is \textit{closed} if
\texttt{y} only consists of \texttt{Zero} and \texttt{Succ}, otherwise we say \texttt{y} is \textit{open}. If \texttt{y : NIncr n a} is open, then it must be of the form $\texttt{Succ}^n\ \texttt{x}$, where \texttt{x : a}.
The behavior of \texttt{mapIncr} is subtle. 
Suppose \texttt{y : NIncr n a} is open, then the result of \texttt{mapIncr l Succ y} (where $l \leq n$) will be \texttt{Succ y}. If \texttt{y : NIncr n a} is closed, then
\texttt{y} must be of the form $\texttt{Succ}^m \ \texttt{Zero}$, where $m < n$. In this case \texttt{mapIncr l Succ y} will be evaluated to \texttt{y} if $m < l \leq n$, or evaluated to \texttt{Succ y} if $l \leq m$. We can test this by comparing
the following \texttt{num1} and \texttt{num2}. The program \texttt{num1} is evaluated to \texttt{Succ (Succ Zero)}, while \texttt{num2} is evaluated to \texttt{Succ (Succ (Succ Zero))}. %% When using \texttt{mapIncr} on \texttt{y : NIncr n a},  most of the time we just need \texttt{mapIncr n f y}, but in some special cases, we want
%% to use \texttt{mapIncr m f y}, where \texttt{m} is strictly smaller than \texttt{n}.

\begin{verbatim}
num0 : NIncr (S (S (S (S (S Z))))) Char
num0 = Succ (Succ Zero)
num1 : NIncr (S (S (S (S (S (S Z)))))) Char
num1 = mapIncr {Incr (Incr Char)} {Incr (Incr (Incr Char))} 
       (S (S (S Z))) Succ num0
num2 : NIncr (S (S (S (S (S (S Z)))))) Char
num2 = mapIncr {Incr (Incr (Incr Char))} {Incr (Incr (Incr (Incr Char)))} 
       (S (S Z)) Succ num0
\end{verbatim}

We now define the following dependently typed fold for $\texttt{Term}\ (\texttt{Incr}^n\ a)$.

\begin{verbatim}
foldT : {a : Set} -> {p : Nat -> Set} -> (n : Nat) ->
        ((m : Nat) -> NIncr m a -> p m) ->
        ((m : Nat) -> p m -> p m -> p m) ->
        ((m : Nat) -> p (S m) -> p m) -> Term (NIncr n a) -> p n
foldT {a} {p} n var app lam (Var x) = var n x
foldT {a} {p} n var app lam (App x1 x2) =
   app n (foldT {a} {p} n var app lam x1) (foldT {a} {p} n var app lam x2) 
foldT {a} {p} n var app lam (Lam x) = lam n (foldT {a} {p} (S n) var app lam x)
\end{verbatim}
The function \texttt{foldT} traverses the structure of \texttt{Term}, replaces the constructors \texttt{Var}, \texttt{App} and \texttt{Lam} with \texttt{var}, \texttt{app} and \texttt{lam}, increases the number \texttt{n} only when traversing under the \texttt{Lam} constructor.

The higher-order fold for \texttt{Term} is an instance of the dependently typed fold \texttt{foldT}.

\begin{verbatim}
hfoldT : {a : Set} -> {p : Set -> Set} -> 
         ({a : Set} -> p a) ->
         ({a : Set} -> p a -> p a -> p a) ->
         ({a : Set} -> p (Incr a) -> p a) -> Term a -> p a
hfoldT {a} {p} var app lam =
  foldT {a} {\ n -> p (NTimes Incr n a)} Z (\ m a -> var) (\ m -> app) (\ m -> lam)   
\end{verbatim}

The following is the map function defined from \texttt{foldT} and \texttt{mapIncr}.

\begin{verbatim}
mapT : {a b : Set} -> (n : Nat) -> (a -> b) -> Term (NIncr n a) -> Term (NIncr n b)
mapT {a} {b} n f y =
  foldT {a} {\ n -> Term (NIncr n b)} n
    (\ n v -> Var (mapIncr n f v)) (\ n -> App) (\ n -> Lam) y
\end{verbatim}
%% Note that \texttt{mapT \string{a\string} \string{Incr a\string} Z f} maps \texttt{Term a} to \texttt{Term b}.
When defining the function \texttt{mapT}, we use \texttt{foldT} to traverse the
abstract data type \texttt{Term (NIncr n a)}, and perform action at the leaves, where
we apply the \texttt{mapIncr} function.

For a variable \texttt{Var x}, we say it is \textit{open}/\textit{closed} iff \texttt{x} is \textit{open}/\textit{closed}.
Note that an open variable is necessarily a free variable, but
a closed variable can be either bound or free. For example, the variable \texttt{Var Zero : Term (Incr Char)}
 is a free and closed variable because it is not bound by any \texttt{Lam} constructor, while the variable \texttt{Var Zero : Term (Incr Char)} in \texttt{Lam (Var Zero) : Term Char} is closed and bound. Let \texttt{y} be a term of type \texttt{Term (NIncr n a)} for some fixed constant type \texttt{a}. The function call \texttt{mapT n f y} will map \texttt{f} to all the open variables in \texttt{y} and leave the closed variables (including free and closed variables) unchanged. On the other hand, the function call \texttt{mapT Z f y} will map the function \texttt{f} to all the free variables (open or closed) in \texttt{y}, while leaving the bound variables unchanged. When $\texttt{n} = \texttt{Z}$ and \texttt{y : Term (Incr Z a)}, there are
no closed free variables in \texttt{y}, thus the free variables coincide with the open variables, the bound variables coincide with the closed variables. 

\subsection{Programming with dependently typed folds and maps}

The following programs are the printing functions
defined from \texttt{foldT} and \texttt{foldI}.

\begin{verbatim}
showT : Term String -> String
showT y = foldT {String} {\ n -> String} Z
                showI
                (\ m x y -> ` LP ` ++ x ++ ` EMP ` ++ y ++ ` RP `) 
                (\ m x -> ` L ` ++ x) y
showI : (m : Nat) -> NIncr m String -> String
showI m y = foldI {String} {\ n -> String} m 
            (\ x -> x) (\ m -> ` Ze `) (\ m x -> ` Su ` ++ x) y
showTC : Term Char -> String
showTC x = showT (mapT Z (\ x -> Cons x Nil) x)
\end{verbatim}
In the definition of \texttt{showTC}, we first convert all the free variables in the term \texttt{x} to \texttt{String}\footnote{The data types \texttt{String} and \texttt{Char} are user-defined.}, and then apply \texttt{showT} to the resulting term. 

We now define the abstraction function \texttt{abst} that takes a name $x$ and a term $t$, returns the abstracted term $\lambda x.t$. We assume there is a generic comparison function\footnote{The comparison function here is similar to the equality method in the \texttt{Eq} type class in Haskell.} \texttt{cmp : \string{a : Set\string} -> a -> a -> Bool} and it satisfies \texttt{axiom1 : \string{a : Set\string} -> (x x1 : a) -> cmp x x1 == True -> x == x1}. We can express these two assumptions as postulates in Agda.

\begin{verbatim}
abst : {a : Set} -> a -> Term a -> Term a
abst x t = Lam (mapT Z (match x) t)
match : {a : Set} -> a -> a -> Incr a
match {a} a1 a2 = foldBool {Incr a} Zero (Succ a2) (cmp a1 a2)
foldBool : {p : Set} -> p -> p -> Bool -> p 
foldBool x y True = x
foldBool x y False = y
\end{verbatim}

The function \texttt{match} compares \texttt{a1} with \texttt{a2}, if they are equal,
returns \texttt{Zero}, else returns \texttt{Succ a2}. So the function call \texttt{abst x t} will
look at all the free variables in \texttt{t}, replace a free variable to \texttt{Zero}
if it is equal to \texttt{x}, otherwise add an extra \texttt{Succ} to it. 

A beta-redex is a term of the form \texttt{App (Lam t) s : Term a}. To perform a beta reduction for \texttt{App (Lam t) s}, we need to substitute the variables bound by \texttt{Lam} in \texttt{t} with \texttt{s}. The following actions are required to define the subsitution. (1) For each variable bound
by the \texttt{Lam} inside \texttt{t}, we need to replace it by \texttt{s}. (2) For each free variable in \texttt{t}, we need to decrease a \texttt{Succ} in it. (3) For each free variable in \texttt{s}, we need to increase a \texttt{Succ} for it whenever \texttt{s} is traversing under a binder in \texttt{t}. (4) All the other bound variables in \texttt{t} remain unchanged.

In the following we define the substitution function \texttt{subst} and the function \texttt{redex}.

\begin{verbatim}
redex : {a : Set} -> Term a -> Term a
redex (App (Lam t) s) = subst Z s t
redex t = t

subst : {a : Set} -> (n : Nat) -> Term a -> 
        Term (NIncr n (Incr a)) -> Term (NIncr n a)
subst {a} n s t =
  foldT {Incr a} {\ n -> Term (NIncr n a)} n
       (\ m -> varcase m s) (\ m -> App) (\ m -> Lam) t

varcase : {a : Set} -> (n : Nat) -> Term a -> NIncr n (Incr a) -> Term (NIncr n a)
varcase {a} n s v =
  foldI {Incr a} {\ n -> Term (NIncr n a)} n
   (h s) 
   (\ m -> Var Zero)               -- Action (4)
   (\ m r -> mapT Z Succ r)        -- Action (3)
   v
  where h : {a : Set} -> Term a -> Incr a -> Term a
        h s Zero = s               -- Action (1) 
        h s (Succ b) = Var b       -- Action (2)
\end{verbatim}
The substitution function \texttt{subst} is generalized to allow substitution for any term
of the type \texttt{Term (NIncr n (Incr a))}, the definition of \texttt{redex} uses \texttt{subst Z}. In the definition of \texttt{subst}, all the actions happen in the variable case, we use \texttt{foldT} just for traversing to the leaves of \texttt{Term (NIncr n (Incr a))}. The \texttt{varcase} function inspects on the variable \texttt{v} and performs actions (1)--(4).

\subsection{Reasoning with the induction principles \texttt{indI} and \texttt{indT}}

\noindent Based on the dependently typed fold \texttt{foldT} and \texttt{foldI}, we can obtain the following
induction principles. The induction principles follow the same definitions as
\texttt{foldT} and \texttt{foldI}, except their types are more general. They provide
a way to prove a property holds for any \texttt{Term (NIncr n a)} and \texttt{NIncr n a}.

\begin{verbatim}
indT : {a : Set} -> {p : (n : Nat) -> Term (NIncr n a) -> Set} ->
       (n : Nat) -> ((m : Nat) -> (v : NIncr m a) -> p m (Var v)) ->
       ((m : Nat) -> {v1 v2 : Term (NIncr m a)} -> 
              p m v1 -> p m v2 -> p m (App v1 v2)) ->
       ((m : Nat) -> {v : Term (NIncr (S m) a)} -> p (S m) v -> p m (Lam v)) ->
       (v : Term (NIncr n a)) -> p n v

indI : {a : Set} -> {p : (n : Nat) -> NIncr n a -> Set} -> (n : Nat) ->
       ((x : a) -> p Z x) ->
       ((m : Nat) -> p (S m) Zero) ->
       ((m : Nat) -> {x : NIncr m a} -> p m x -> p (S m) (Succ x)) -> 
       (v : NIncr n a) -> p n v
\end{verbatim}

Now we can use induction to prove the following \texttt{thm1}, which states $(\lambda x . t)\ x =_\beta t$. 

\begin{verbatim}
thm1 : {a : Set} -> (x : a) -> (t : Term a) -> 
         redex (App (abst x t) (Var x)) == t
thm1 {a} x t =
    indT {a} {\ n v -> subst n (Var x) (mapT n (match x) v) == v} Z
      (thm0 x) 
      (\ m {v1} {v2} ih1 ih2 -> cong2 App ih1 ih2) 
      (\ m {v} ih -> cong Lam ih) t

thm0 : {a : Set} -> (x : a) -> (m : Nat) -> (v : NIncr m a) ->
         subst m (Var x) (mapT m (match x) (Var v)) == Var v
thm0 {a} x m v =
    indI {a} {\ n u -> subst n (Var x) (mapT n (match x) (Var u)) == Var u} m
      (\ x1 -> lemm0 (cmp x x1) (lemm1 x x1) (lemm2 x x1))
      (\ m -> refl) (\ m {x1} ih -> cong (mapT Z Succ) ih) v

lemm0 : {p : Set} -> (x : Bool) -> (x == True -> p) -> (x == False -> p) -> p
lemm1 : {a : Set} -> (x x1 : a) -> (cmp x x1 == True) ->
        subst Z (Var x) (mapT {a} {Incr a} Z (match {a} x) (Var x1)) == Var x1
lemm2 : {a : Set} -> (x x1 : a) -> (cmp x x1 == False) ->
        subst Z (Var x) (mapT {a} {Incr a} Z (match {a} x) (Var x1)) == Var x1
\end{verbatim}
The inductive cases for \texttt{thm1} are straightforward, the only nontrivial
case is the variable case, which needs \texttt{thm0}. For \texttt{thm0}, the only nontrivial
case is the base case, where we need a proof of \texttt{(x x1 : a) -> subst Z (Var x) (mapT Z (match x) (Var x1)) == Var x1}. We prove this by considering both \texttt{cmp x x1 == True} (\texttt{lemm1}) and \texttt{cmp x x1 == False} (\texttt{lemm2}). The proofs of \texttt{lemm1} (requires \texttt{axiom1}), \texttt{lemm2} and \texttt{lemm0} are straightforward.

We now consider the theorems \texttt{mapTfuseZ} and \texttt{mapSubst}, which are properties of \texttt{mapT} and \texttt{subst} and will be needed for proving the theorem \texttt{cvtThm} in the Section \ref{terme}.

The following \texttt{mapTfuseZ} theorem states that for any \texttt{s : Term (NIncr n a)},
the following two results are equal: (1) first add an extra \texttt{Succ} to
all the free variables, and then add an extra \texttt{Succ} to all the open variables.
(2) first add an extra \texttt{Succ} to all the open variables, and then add an extra \texttt{Succ} to all the free variables. 

\begin{verbatim}
mapTfuseZ : {a : Set} -> (n : Nat) -> (s : Term (NIncr n a)) ->
            mapT {a} {Incr a} (S n) Succ 
                 (mapT {NIncr n a} {Incr (NIncr n a)} Z Succ s) ==
            mapT {Incr (NIncr n a)} {Incr (Incr (NIncr n a))} Z Succ
                 (mapT {a} {Incr a} n Succ s)
mapTfuseZ {a} n s = mapTfuse {a} Z n s        

mapTfuse : {a : Set} -> (m n : Nat) -> (s : Term (NIncr (add m n) a)) ->
            mapT {a} {Incr a} (S (add m n)) Succ
              (mapT {NIncr n a} {Incr (NIncr n a)} m Succ s) ==
            mapT {NIncr (S n) a} {Incr (NIncr (S n) a)} m Succ
              (mapT {a} {Incr a} (add m n) Succ s)
\end{verbatim}
The \texttt{mapTfuseZ} theorem requires us to prove a more general
lemma \texttt{mapTfuse}, finding this lemma takes a lot of effort, but it can be proved by standard induction using \texttt{indT} and \texttt{indI}. 

The following \texttt{mapSubst} theorem states how \texttt{mapT Z Succ} commutes with \texttt{subst m}. Again, \texttt{mapSubst} requires a nontrivial general lemma \texttt{mapSubst'}. The lemma
\texttt{mapSubst'} can be proved by induction using the lemma \texttt{mapTfuseZ}. %%  in the Section
%% \ref{term}.

\begin{verbatim}
mapSubst : {a : Set} -> (m : Nat) -> (s : Term a) -> (t : Term (NIncr (S m) a)) ->
           mapT Z Succ (subst m s t) == subst (S m) s (mapT Z Succ t)
mapSubst {a} m s t = mapSubst' {a} Z m s t            
mapSubst' : {a : Set} -> (n m : Nat) -> (s : Term a) ->
            (t : Term (NIncr (S (add n m)) a)) ->
            mapT {NIncr m a} {NIncr (S m) a} n Succ (subst (add n m) s t) ==
            subst (S (add n m)) s 
               (mapT {NIncr (S m) a} {NIncr (S (S m)) a} n Succ t)
\end{verbatim}

\section{Case study II: de Bruijn notation as the nested data type \texttt{TermE}}
\label{terme}
In this section we consider representing the de Bruijn lambda terms using the following nested data type.

\begin{verbatim}
data TermE (a : Set) : Set where
  VarE : a -> TermE a
  AppE : TermE a -> TermE a -> TermE a
  LamE : TermE (Incr (TermE a)) -> TermE a
\end{verbatim}
This data type is already motivated by Bird and Paterson \cite{bird1999bruijn}. Recall that
when substituting term $s$ for $x$ in term $t$, i.e. $[s/x]t$, 
we need to perform the actions (1)--(4) (Section~\ref{term}). The action (3) requires traversing the term $s$ to
add an additional \texttt{Succ} when the substitution is going under a binder. This
has two drawbacks, namely, traversing $s$ takes additional time and
prevents the sharing of $s$.

For example, the redex $(\lambda . 0 \ (\lambda . 1 \ 0 \ (\lambda . 2 \ 1 \ 0))) \ (\lambda . 0 \ (S \ 'W'))$ will be reduced to the following.

\begin{center}
  $\underline{(\lambda . 0 \ (S \ 'W'))} \ (\lambda . \underline{(\lambda . 0 \ (S (S \ 'W')))} \ 0 \ (\lambda . \underline{(\lambda . 0 \ (S (S (S \ 'W'))))} \ 1 \ 0))$
\end{center}

\noindent Not only we traverse the term $(\lambda . 0 \ (S \ 'W'))$ three times to add $S$, but also we have three different copies of $(\lambda . 0 \ (S \ 'W'))$ in the resulting term. A more efficient implementation
 would avoid such traversal and allow us to obtain the following term. 

 \begin{center}
 $\texttt{term0} = \underline{(\lambda . 0 \ (S \ 'W'))} \ (\lambda . (S \underline{(\lambda . 0 \ (S \ 'W'))}) \ 0 \ (\lambda . (S (S \underline{(\lambda . 0 \ (S \ 'W'))})) \ 1 \ 0))$
\end{center}

 \noindent In \texttt{term0}, we have three same copies of $(\lambda . 0 \ (S \ 'W'))$.
 To enable such representation, we would need to allow
  \texttt{Succ} to be applied to a term, hence the type for the \texttt{LamE} constructor.
 The following \texttt{term1} is the concrete representation of $(\lambda . 0 \ (S \ 'W'))$ and \texttt{term2} is the representation of \texttt{term0}.

\begin{verbatim}
term1 : TermE Char
term1 = LamE (AppE (VarE Zero) (VarE (Succ (VarE W))))

term2 : TermE Char
term2 = AppE term1
             (LamE (AppE (AppE (VarE (Succ term1))
                               (VarE Zero))
                         (LamE (AppE (AppE (VarE (Succ (VarE (Succ term1))))
                                           (VarE (Succ (VarE Zero))))
                                     (VarE Zero)))))
\end{verbatim}

\subsection{The dependently typed fold \texttt{foldE}}

At each level down the constructor \texttt{LamE}, a term gains an additional type-constructor \texttt{\string\ x -> Incr (TermE x)} in its type. So we will focus on the type of the form $\texttt{TermE}\ (\texttt{IncrTermE}^n\ a)$, where $\texttt{IncrTermE}^n$ means the $n$-th iteration of $\texttt{\string\ x -> Incr (TermE x)}$.

We define the following dependently typed fold \texttt{foldE} for $\texttt{TermE}\ (\texttt{IncrTermE}^n\ a)$. 

\begin{verbatim}
IncrTermE : Nat -> Set -> Set
IncrTermE = NTimes (\ x -> Incr (TermE x))

foldE : {a : Set} -> {p : Nat -> Set} -> (n : Nat) ->
         (a -> p Z) ->
         ((m : Nat) -> p (S m)) ->
         ((m : Nat) -> p m -> p (S m)) -> 
         ((m : Nat) -> p m -> p m -> p m) ->
         ((m : Nat) -> p (S m) -> p m) -> TermE (IncrTermE n a) -> p n 
foldE {a} {p} Z varBase varZero varSucc app abs (VarE x) = varBase x
foldE {a} {p} (S n) varBase varZero varSucc app abs (VarE Zero) = varZero n
foldE {a} {p} (S n) varBase varZero varSucc app abs (VarE (Succ x)) =
  varSucc n (foldE {a} {p} n varBase varZero varSucc app abs x)
foldE {a} {p} n varBase varZero varSucc app lam (LamE x) =
  lam n (foldE {a} {p} (S n) varBase varZero varSucc app lam x)
foldE {a} {p} n varBase varZero varSucc app abs (AppE x x') =
  app n (foldE {a} {p} n varBase varZero varSucc app abs x)
        (foldE {a} {p} n varBase varZero varSucc app abs x')
\end{verbatim}

Observe that \texttt{foldE} is well-founded and the abstract indexed data type $\texttt{TermE}\ (\texttt{IncrTermE}^n\ a)$ 
has five constructors, corresponding to the five cases in the definition of \texttt{foldE}.
%% There are three sub-cases for the variable, they corresponds to the variable of type
%%  $\texttt{TermE}\ a$ or $\texttt{TermE}\ (\texttt{IncrTermE}^{n+1}\ a)$, where $n \geq 0$. 
There are two variable cases for $\texttt{TermE}\ (\texttt{IncrTermE}^{n+1}\ a)$, i.e. $\texttt{VarE Zero}$ and $\texttt{VarE (Succ x)}$, where \texttt{x} is of
the type $\texttt{TermE}\ (\texttt{IncrTermE}^{n}\ a)$.
So there is a recursive
call in the case for \texttt{VarE (Succ x)}, which traverses the term \texttt{x}.

The following is the map function defined from \texttt{foldE}.

\begin{verbatim}
mapE : {a b : Set} -> (n : Nat) -> (a -> b) -> 
       TermE (IncrTermE n a) -> TermE (IncrTermE n b)
mapE {a} {b} n f x = foldE {a} {\ n -> TermE (IncrTermE n b)} n
                      (\ x -> VarE (f x))
                      (\ m -> VarE Zero)
                      (\ m r -> VarE (Succ r))
                      (\ m -> AppE)
                      (\ m -> LamE) x
\end{verbatim}

\noindent The function call \texttt{mapE n f t} keeps every constructor in \texttt{t} unchanged and only applies the function \texttt{f} in the case of \texttt{VarE x}, where \texttt{x : a}. The map function for \texttt{TermE a} is \texttt{mapE Z}.

The dependently typed fold \texttt{foldE} can be specialized to the higher-order fold for \texttt{TermE}.

\begin{verbatim}
hfoldE : {a : Set} -> {p : Set -> Set} ->
          ({a : Set} -> a -> p a) ->
          ({a : Set} -> p a -> p a -> p a) ->
          ({a : Set} -> p (Incr (p a)) -> p a) -> TermE a -> p a
hfoldE {a} {p} var app lam =
  foldE {a} {\ n -> p (NTimes (\ y -> Incr (p y)) n a)}
    Z var (\ m -> var Zero) (\ m r -> var (Succ r)) (\ m -> app) (\ m -> lam)
\end{verbatim}
The higher-order fold \texttt{hfoldE} replaces the constructors
\texttt{VarE}, \texttt{AppE} and \texttt{LamE} to \texttt{var}, \texttt{app} and \texttt{lam}
while keeping the constructors \texttt{Zero} and \texttt{Succ} unchanged.  

\subsection{Programming with \texttt{foldE}}

\noindent The following function \texttt{redexE} is for reducing a beta-redex.
The substitution function \texttt{substE} follows the same idea as the \texttt{subst} function
in Section \ref{term}, except that it does not perform action (3) due to the
optimization of the \texttt{TermE} data type.

\begin{verbatim}
redexE : {a : Set} -> TermE a -> TermE a
redexE (AppE (LamE t) s) = substE Z s t
redexE t = t

substE : {a : Set} -> (n : Nat) -> TermE a -> 
         TermE (IncrTermE n (Incr (TermE a))) -> TermE (IncrTermE n a)
substE {a} n s = foldE {Incr (TermE a)} {\ n -> TermE (IncrTermE n a)} n
                  (base s) 
                  (\ m -> VarE Zero)        -- Action (4)
                  (\ m r -> VarE (Succ r))  -- No need for action (3)
                  (\ m -> AppE) 
                  (\ m -> LamE)
  where base : TermE a -> Incr (TermE a) -> TermE a
        base s Zero = s                     -- Action (1)
        base s (Succ x) = x                 -- Action (2)
\end{verbatim}

%% Although the data type \texttt{TermE} is more complicated than \texttt{Term}, but the
%% beta-reduction function \texttt{redexE} and its verification is actually
%% simpler than \texttt{redex}. 
The following abstraction function \texttt{abstE x t} binds
the free variable \texttt{x} in \texttt{t}. The definition of \texttt{abstE} follows the
same idea as \texttt{abst} (Section \ref{term}). %% and the \texttt{cmp} is the comparison function with the same postulation \texttt{axiom1} we used in Section \ref{term}.

\begin{verbatim}
abstE : {a : Set} -> a -> TermE a -> TermE a
abstE x t = LamE (mapE Z (matchE x) t)
matchE : {a : Set} -> a -> a -> Incr (TermE a)
matchE {a} a1 a2 = foldBool {Incr (TermE a)} Zero (Succ (VarE a2)) (cmp a1 a2)
\end{verbatim}

Let us now consider converting a lambda expression of the type \texttt{TermE a} to a lambda
expression of the type \texttt{Term a}. We will define a generalized conversion function
to convert \texttt{TermE (IncrTermE n a)} to \texttt{Term (NIncr n a)} for any \texttt{n}.    
The main difference between \texttt{TermE (IncrTermE n a)} and \texttt{Term (NIncr n a)}
is that \texttt{Succ} can apply to a term \texttt{t : TermE (IncrTermE n a)}, so there are terms of the form \texttt{VarE (Succ t) : TermE (IncrTermE (S n) a)}. The key idea of the conversion
 is that when converting a term of
 the form \texttt{VarE (Succ t)}, we first convert the term \texttt{t} of type \texttt{TermE (IncrTermE n a)} to a term \texttt{t'} of type \texttt{Term (NIncr n a)}, and then apply the constructor \texttt{Succ} to all the free variables in \texttt{t'}, i.e. \texttt{mapT Z Succ t'}. %The following is the conversion function. 

%% Note that here by the free variables in \texttt{t'}, we means the variable that are not bound in \texttt{t'}, these include variables
%% of the forms \texttt{Var x} (for some \texttt{x} of type \texttt{a}) and \texttt{Var Zero}. So
%% it is not sufficient to write \texttt{mapT n Succ t'}, as this will only apply \texttt{Succ}
%% to free variables of the forms \texttt{Var x}. To also apply \texttt{Succ} to \texttt{Var Zero},
%% we need \texttt{mapT Z Succ t'}. Because the free variable \texttt{Var Zero} will be of the type \texttt{Term (NIncr (S n) a)}, to obtain \texttt{Var (Succ Zero)}, we cannot use \texttt{mapT (S n) Succ (Var Zero)}, but \texttt{mapT m Succ (Var Zero)}, where \texttt{m} has to be strictly
%% smaller than \texttt{S n}, to also account for \texttt{Var x} of type \texttt{Term (NIncr Z a)},
%% we can only use \texttt{m = Z}. Hence we have the following conversion function.  

\begin{verbatim}
cvtE : {a : Set} -> (n : Nat) -> TermE (IncrTermE n a) -> Term (NIncr n a)   
cvtE {a} n t = foldE {a} {\ n -> Term (NIncr n a)} n
                (\ a -> Var a) 
                (\ m -> Var Zero)
                (\ m t' -> mapT Z Succ t')
                (\ m -> App) 
                (\ m -> Lam) t
\end{verbatim}

The function \texttt{cvtE} converts \texttt{VarE x}, \texttt{VarE Zero}, \texttt{AppE} and \texttt{LamE} to \texttt{Var x}, \texttt{Var Zero}, \texttt{App} and \texttt{Lam} accordingly. The
only subtle case is how to convert \texttt{VarE (Succ t)}, which we have explained.

\subsection{Reasoning with the induction principle \texttt{indE}}

\noindent The following is the induction principle for \texttt{TermE (IncrTermE n a)}, we elide the definition as it is the same as \texttt{foldE}.

\begin{verbatim}
indE : {a : Set} -> {p : (n : Nat) -> TermE (IncrTermE n a) -> Set} -> (n : Nat) -> 
       ((x : a) -> p Z (VarE x)) ->
       ((m : Nat) -> p (S m) (VarE Zero)) ->
       ((m : Nat) -> {r : TermE (IncrTermE m a)} -> 
                 p m r -> p (S m) (VarE (Succ r))) ->
       ((m : Nat) -> {x1 x2 : TermE (IncrTermE m a)} ->
                 p m x1 -> p m x2 -> p m (AppE x1 x2)) ->
       ((m : Nat) -> {x : TermE (IncrTermE (S m) a)} ->
                 p (S m) x -> p m (LamE x)) -> 
       (v : TermE (IncrTermE n a)) -> p n v
\end{verbatim}

The induction principle \texttt{indE} gives us a way to prove a property \texttt{p : (n : Nat) -> TermE (IncrTermE n a) -> Set} holds for any \texttt{v : TermE (IncrTermE n a)}. To obtain such proof, we first have to prove \texttt{p} holds for \texttt{VarE x : TermE (IncrTermE Z a)} and \texttt{VarE Zero : TermE (IncrTermE (S m) a)}. They correspond to the two arguments \texttt{(x : a) -> p Z (VarE x)}
and \texttt{(m : Nat) -> p (S m) (VarE Zero)} for \texttt{indE}. Then we will need to prove
three inductive cases, they correspond to the other three arguments for \texttt{indE}. For example,
one inductive case requires us to prove \texttt{p m (LamE x)}, using the inductive hypothesis \texttt{p (S m) x}. %, where \texttt{x : TermE (IncrTermE (S m) a)}.

We now use \texttt{indE} to prove that $(\lambda x.t)\ x =_\beta t$ holds for our definition of \texttt{redexE}. The proof is similar to the proof of theorem \texttt{thm1} for \texttt{redex} in Section \ref{term}.

\begin{verbatim}
thm1 : {a : Set} -> (x : a) -> (t : TermE a) -> 
       redexE (AppE (abstE x t) (VarE x)) == t
thm1 {a} x t = 
  indE {a} {\ n v -> substE n (VarE x) (mapE n (matchE x) v) == v} Z
      (lem3 x) (\ m -> refl) (\ m {r} ih -> cong (\ y -> VarE (Succ y)) ih)
      (\ m ih1 ih2 -> cong2 AppE ih1 ih2) (\ m ih -> cong LamE ih) t
lem3 : {a : Set} -> (x y : a) -> 
         substE Z (VarE x) (mapE Z (matchE x) (VarE y)) == VarE y
\end{verbatim}

To convince ourself that the conversion function \texttt{cvtE} is well-behaved, we use \texttt{indE} prove
that the conversion function commutes with the substitution, i.e. $\texttt{cvtE}([s/x]t) = [(\texttt{cvtE}\ s)/x](\texttt{cvtE}\ t)$.
%% if we convert the result of substituting \texttt{s} in \texttt{t}, it is the same as we first convert both \texttt{s} and \texttt{t} and then perform the substitution.
{
\begin{verbatim}
cvtThm : {a : Set} -> (n : Nat) -> (s : TermE a) ->
         (t : TermE (IncrTermE (S n) a)) ->
         cvtE {a} n (substE n s t) == subst n (cvtE {a} Z s) (cvtE {a} (S n) t)
cvtThm {a} n s t =
  indE {Incr (TermE a)} 
       {\ n v -> cvtE {a} n (substE n s v) == 
                 subst n (cvtE {a} Z s) (cvtE {a} (S n) v)} n
       (\ x -> lemmVar x s)   -- case VarE x
       (\ m -> refl)          -- case VarE Zero
       (\ m {r} ih ->         -- case VarE (Succ r)
        equational cvtE {a} (S m) (substE (S m) s (VarE (Succ r)))
          equals cvtE {a} (S m) (VarE (Succ (substE m s r))) by refl
          equals mapT {NIncr m a} {Incr (NIncr m a)} Z Succ 
                      (cvtE {a} m (substE m s r)) 
             by refl
          equals mapT {NIncr m a} {Incr (NIncr m a)} Z Succ
                      (subst m (cvtE {a} Z s) (cvtE {a} (S m) r))
             by cong (mapT {NIncr m a} {Incr (NIncr m a)} Z Succ) ih
          equals subst (S m) (cvtE {a} Z s)
                  (mapT {NIncr (S m) a} {Incr (Incr (NIncr m a))} Z Succ
                        (cvtE {a} (S m) r))
             by mapSubst {a} m (cvtE {a} Z s) (cvtE {a} (S m) r)
          equals subst (S m) (cvtE {a} Z s) 
                       (cvtE {a} (S (S m)) (VarE (Succ r))) 
             by refl)
       (\ m {x1} {x2} ih1 ih2 -> cong2 App ih1 ih2) -- case AppE x1 x2
       (\ m {x} ih -> cong Lam ih)                  -- case LamE x
       t
\end{verbatim}
}
%% In the proof of \texttt{cvtThm}, 

The cases for \texttt{LamE x}, \texttt{AppE x1 x2} and \texttt{VarE Zero} are straightforward.
For the case of \texttt{VarE x}, we need the following \texttt{lemmVar} and \texttt{lemmM}.

\begin{verbatim}
lemmVar : {a : Set} -> (x : Incr (TermE a)) -> (s : TermE a) ->
           cvtE Z (substE Z s (VarE x)) == subst Z (cvtE Z s) (cvtE (S Z) (VarE x))
lemmVar {a} Zero s = refl
lemmVar {a} (Succ x) s = lemmM {a} Z (cvtE Z s) (cvtE Z x)

lemmM : {a : Set} -> (m : Nat) -> (s : Term a) -> (t : Term (NIncr m a)) ->
        t == subst m s (mapT {a} {Incr a} m Succ t)
\end{verbatim}
  %% indT {a} {\ n v -> v == subst n s (mapT {a} {Incr a} n Succ v)} m
  %% (\ m v -> indI {a} {\ n t -> Var t == subst n s (mapT {a} {Incr a} n Succ (Var t))} 
  %%             m (\ x -> refl) (\ m -> refl) (\ x ih -> cong (mapT Z Succ) ih) v)
  %% (\ m ih1 ih2 -> cong2 App ih1 ih2) (\ m {v} ih -> cong Lam ih) x                

\noindent The lemma \texttt{lemmVar} is just a special case of \texttt{cvtThm} for \texttt{VarE x}, it needs the lemma \texttt{lemmM}, which can be proved by straightforward induction.

For the case of \texttt{VarE (Succ r)}, 
we need the induction hypothesis \texttt{ih} and the \texttt{mapSubst} theorem we proved in Section \ref{term}. We also use a custom equational reasoning tactic in
the following form \footnote{Please see the file \texttt{Equality.agda} in the supplementary material for details.}.

\begin{verbatim}
equational t1 
  equals t2 by p1 
  equals t3 by p2 ...
  equals tn by p(n-1)
\end{verbatim}
This means we prove the following $t_1 \stackrel{p_1}{=} t_2 \stackrel{p_2}{=} t_3 ... \stackrel{p_{n-1}}{=} t_n$.

\subsection{Discussion}
%% We have shown how to program and reason about dependently typed folds using the nested data type \texttt{Bush}, \texttt{Term} and \texttt{TermE} as examples.
%% Programming with nested data types is challenging, but dependently typed folds can help in that they enable a form of structural programming, and provide us with the induction principles so that formal verification is possible.
We have shown that not only we are able to program all the functions for manipulating de Bruijn  lambda terms like Bird and Paterson did \cite{bird1999bruijn},
but also we are able to reason about the programs formally using induction principles. We are working in a terminating dependently typed language, whereas Bird
and Paterson worked in Haskell, a language with general recursion.   
%% We are working in a terminating dependently typed
%% language, whereas Bird and Paterson worked in Haskell, a language with general recursion.
Benefiting from the flexibility of dependently typed folds, we also make some minor improvements. The following are Bird and Patterson's implementation of \texttt{cvtE} and \texttt{cvtBodyE} functions in Haskell. 
\begin{verbatim}
cvtE :: TermE a -> Term a
cvtE = gfoldE Var App (Lam . joinT . mapT distT) id

cvtBodyE :: TermE (Incr (TermE a)) -> Term (Incr a)
cvtBodyE = joinT . mapT distT . cvtE . mapE (mapI cvtE)
\end{verbatim}
The functions \texttt{gfoldE}, \texttt{joinT}, \texttt{mapE} and \texttt{mapT} all require traversal over a term structure. Bird and Paterson's implementation of \texttt{cvtE}
requires three traversal functions and \texttt{cvtBodyE} requires five traversal
functions. While our implementation of \texttt{cvtE} only requires two traversal functions, i.e. \texttt{foldE} and \texttt{mapT}. Moreover, the function \texttt{cvtE} corresponds to \texttt{cvtE Z} in our implementation and \texttt{cvtBodyE} corresponds to \texttt{cvtE (S Z)}.
So our implementation of \texttt{cvtE} is more flexible.

\section{Obtaining the dependently typed folds for any nested data types}
\label{arbitrary}
In this section, we hint at how the construction of dependently typed fold in Section \ref{dev} can be generalized to arbitrary nested data types. We do not provide a general
formulation of the construction because of the
technical overhead in formulating the most general form of nested data type declarations.
One can find a formulation of nested data types as the fix points of the polynomial higher-order functors in \cite{bird1999generalised}. Instead, we give an example of nested data type that is hopefully general enough
to make it clear what one would do in the general case. We leave the general formulation and the meta-theoretical study of dependently typed folds as future work.

In the previous sections, we use a natural number \texttt{n} as the index
to obtain the dependently typed folds for \texttt{NBush n a}, \texttt{Term (NIncr n a)} and \texttt{TermE (IncrTermE n a)}. While using the natural number index has the benefit of being intuitive, it raises the question of whether it is always possible to obtain dependently typed folds for nested data types. 
%% \footnote{The concept of nested data types we consider in this paper is the one defined by Bird and Paterson \cite[\S 3.2]{bird1999generalised}, it does not include negative data types.}. %% as 
%% so far our approach seems to rely on the natural number index and we only obtain a small number of dependently typed folds for the nested data types from the literature.
Fortunately, in a dependently typed language, types can depend not only on the natural numbers but also on any inductive data types. In a general setting, given an arbitrarily
complicated nested data type, we can use a customized regular data type as the index to define the dependently typed fold.  %% The method we describe in Section \ref{gen} does not work for negative data types, nor data types with higher-kinded type variables.

%% is the strictly positive fragment of the inductive data types \cite{coquand1990inductively}, so it does include GADTs and negative data types.

\subsection{A method to obtain dependently typed folds}
\label{gen}

Consider the following nested data type \texttt{I} and \texttt{D}.

\begin{verbatim}
data I (a : Set) : Set where
  Zero : I a
  Succ : a -> I (I a) -> I a

data D (a b : Set) : Set where
  DNil : D a b
  DCons : a -> b -> D (I a) b -> D (D (I b) (I b)) (I a) -> D a b
  ACons : I b -> D (I (I (D b a))) (D (D b a) (D a b)) -> D a b
\end{verbatim}
Here \texttt{D} is a type-constructor of arity 2 and \texttt{I} is a type-constructor
of arity 1. The nested data type \texttt{D} is reasonably arbitrary and general. It
refers to another nested data type \texttt{I}, and its constructors \texttt{DCons} and \texttt{ACons} are also nested. We will consider how to obtain a dependently typed fold for \texttt{D}.

We first define the following regular data type \texttt{IndexD} to describe all the types arising from \texttt{D}, i.e. the types constructed from \texttt{D}, \texttt{I} and the variables \texttt{a}, \texttt{b}.

\begin{verbatim}
data IndexD : Set where
  VarA : IndexD
  VarB : IndexD
  IsD : IndexD -> IndexD -> IndexD
  IsI : IndexD -> IndexD
\end{verbatim}
The constructors \texttt{VarA} and \texttt{VarB} describe the two variables for \texttt{D}, the constructor \texttt{IsD} of arity 2 describes the type-constructor
\texttt{D} and the constructor \texttt{IsI} of arity 1 describes \texttt{I}.

We then use structural recursion to define the following type-level function 
 that translates a value of \texttt{IndexD} to its corresponding type.

\begin{verbatim}
H : IndexD -> Set -> Set -> Set
H VarA a b = a
H VarB a b = b
H (IsD x y) a b = D (H x a b) (H y a b)
H (IsI x) a b = I (H x a b)
\end{verbatim}
For example, \texttt{H (IsD (IsI (IsD (IsI VarA) (IsI VarB))) (IsI VarA)) Nat Char} will
be evaluated to the type \texttt{D (I (D (I Nat) (I Char))) (I Nat)}. 

Similar to \texttt{NBush n a}, we view \texttt{H i a b} as a kind of
abstract indexed data type. We now define the dependently typed fold for \texttt{H i a b} interactively in Agda. We begin with the following.

\begin{verbatim}
foldD : {a b : Set} {p : IndexD -> Set} -> (i : IndexD) -> H i a b -> p i
foldD {a} {b} {p} i l = ?  
\end{verbatim}
%% Here \texttt{foldD \string{a\string} \string{b\string} \string{p\string} (IsB VarA VarB)} has type \texttt{B a b -> p (IsB VarA VarB)}.  %% and \texttt{foldD \string{a\string} \string{b\string} (IsI VarA)} has type \texttt{I a -> p (IsB VarA)}. So \texttt{foldD} is also a fold function for \texttt{I}.

We will extend the type of \texttt{foldD} with arguments that correspond to the
constructors of the abstract data type \texttt{H i a b}. By dependent pattern-matching on \texttt{i} and \texttt{H i a b}, we have the following seven cases. 

\begin{verbatim}
foldD : {a b : Set} {p : IndexD -> Set} -> (i : IndexD) -> H i a b -> p i
foldD {a} {b} {p} VarA l = ?
foldD {a} {b} {p} VarB l = ?
foldD {a} {b} {p} (IsD i j) DNil = ?
foldD {a} {b} {p} (IsD i j) (DCons x y l v) = ?
foldD {a} {b} {p} (IsD i j) (ACons x l) = ?
foldD {a} {b} {p} (IsI i) Zero = ?
foldD {a} {b} {p} (IsI i) (Succ x y) = ?
\end{verbatim}

The cases for \texttt{VarA}, \texttt{VarB} and \texttt{DNil} suggest that we extend \texttt{foldD} with the arguments \texttt{varA}, \texttt{varB} and \texttt{bnil}.

\begin{verbatim}
foldD : {a b : Set} {p : IndexD -> Set} ->
        (i : IndexD) -> 
        (varA : a -> p VarA) ->
        (varB : b -> p VarB) ->
        (bnil : {i j : IndexD} -> p (IsD i j)) ->
         H i a b -> p i
foldD {a} {b} {p} VarA varA varB bnil l = varA l 
foldD {a} {b} {p} VarB varA varB bnil l = varB l
foldD {a} {b} {p} (IsD i j) varA varB bnil DNil = bnil
foldD {a} {b} {p} (IsD i j) varA varB bnil (DCons x y l v) = ?
...
\end{verbatim}
%% foldD {a} {b} {p} (IsB i j) varA baseB bnil (ACons x l) = ?
%% foldD {a} {b} {p} (IsI i) varA baseB bnil Zero = ?
%% foldD {a} {b} {p} (IsI i) varA baseB bnil (Succ x y) = ?

We will now focus on the case for \texttt{DCons x y l v}, as the other cases
follow similarly. We want to make a recursive call of \texttt{foldD} on each of the components in
\texttt{DCons x y l v}. The following is the environment provided by Agda.

\begin{verbatim}
Goal: p (IsD i j)
-------------------------------------------------------
v     : D (D (I (H j a b)) (I (H j a b))) (I (H i a b))
l     : D (I (H i a b)) (H j a b)
y     : H j a b
x     : H i a b
\end{verbatim}
The key to make these recursive calls is to provide \texttt{foldD} with
the correct indexes and these indexes can be structurally larger than \texttt{IsD i j}.
For example, the index we provide to \texttt{foldD} when calling it on \texttt{v} is
\texttt{IsD (IsD (IsI j) (IsI j)) (IsI i)} and the result of the recursive call is of the type
 \texttt{p (IsD (IsD (IsI j) (IsI j)) (IsI i))}.
 Thus we add the argument \texttt{bcons} for \texttt{foldD} to
 combine the results of the recursive calls on \texttt{x, y, l, v}.

\begin{verbatim}
foldD : {a b : Set} {p : IndexD -> Set} ->
        (i : IndexD) -> 
        (varA : a -> p VarA) ->
        (varB : b -> p VarB) ->
        (bnil : {i j : IndexD} -> p (IsD i j)) ->
        (bcons : {i j : IndexD} -> p i -> p j -> p (IsD (IsI i) j) ->
                p (IsD (IsD (IsI j) (IsI j)) (IsI i)) -> p (IsD i j)) ->
         H i a b -> p i
foldD {a} {b} {p} VarA varA varB bnil bcons l = varA l 
foldD {a} {b} {p} VarB varA varB bnil bcons l = varB l
foldD {a} {b} {p} (IsD i j) varA varB bnil bcons DNil = bnil
foldD {a} {b} {p} (IsD i j) varA varB bnil bcons (DCons x y l v) = 
  bcons
   (foldD {a} {b} {p} i varA varB bnil bcons x)
   (foldD {a} {b} {p} j varA varB bnil bcons y)
   (foldD {a} {b} {p} (IsD (IsI i) j) varA varB bnil bcons l)
   (foldD {a} {b} {p} (IsD (IsD (IsI j) (IsI j)) (IsI i)) 
      varA varB bnil bcons v)
...
\end{verbatim}
Note that the above recursive definition of \texttt{foldD} is well-founded, Agda
is able to confirm its termination. The type of the final definition of \texttt{foldD} is the following, its full definition is in the supplementary material.

\begin{verbatim}
foldD : {a b : Set} {p : IndexD -> Set} ->
       (i : IndexD) ->
       (varA : a -> p VarA) ->
       (varB : b -> p VarB) ->
       (bnil : {i j : IndexD} -> p (IsD i j)) ->
       (bcons : {i j : IndexD} -> p i -> p j -> p (IsD (IsI i) j) ->
                  p (IsD (IsD (IsI j) (IsI j)) (IsI i)) -> p (IsD i j)) ->
       (acons : {i j : IndexD} -> p (IsI j) ->
                  p (IsD (IsI (IsI (IsD j i))) (IsD (IsD j i) (IsD i j))) ->
                  p (IsD i j)) ->          
       (zero : {i : IndexD} -> p (IsI i)) ->
       (succ : {i : IndexD} -> p i -> p (IsI (IsI i)) -> p (IsI i)) ->
       H i a b -> p i
\end{verbatim}

We can use the dependently typed fold \texttt{foldD} to define the following
map and sum functions. 

\begin{verbatim}
mapD : {a b c d : Set} -> (i : IndexD) -> (a -> c) -> (b -> d) -> H i a b -> H i c d
mapD {a} {b} {c} {d} i f g l = 
  foldD {a} {b} {\ i -> H i c d} i f g DNil DCons ACons Zero Succ l 

mapD' : {a b c d : Set} -> (a -> c) -> (b -> d) -> D a b -> D c d
mapD' = mapD (IsD VarA VarB)

sumD : D Nat Nat -> Nat
sumD x =  foldD {Nat} {Nat} {\ i -> Nat} (IsD VarA VarB) (\ y -> y) (\ y -> y)
  Z (\ x1 x2 x3 x4 -> add x1 (add x2 (add x3 x4))) add Z add x
\end{verbatim}
%% mapD'' : {a c : Set} -> (a -> c) -> I a -> I c
%% mapD'' {a} {c} f = mapD {a} {a} {c} {c} (IsI VarA) f f
The map function \texttt{mapD} traverses over the abstract data type \texttt{H i a b}, leaving all the
constructors unchanged, while applying \texttt{f} and \texttt{g} at the leaves. 

Similar to what we have described in the previous sections, we can obtain an induction principle
from \texttt{foldD} by generalizing its type, i.e., generalizing the kind of \texttt{p}
 to \texttt{(i : IndexD) -> H i a b -> Set}. Moreover, since \texttt{H i a b} 
can be viewed as an indexed data type with seven constructors, we can obtain the indexed representation (Section \ref{index}) and the Church-encoding of \texttt{H} (Section \ref{church}). Finally, it should be clear that we can apply the method
we just described to obtain dependently typed folds for any nested data types. 

\subsection{Specializing dependently typed folds to higher-order folds}
\label{well-formed}
Let us consider how to specialize dependently typed folds to the higher-order folds, using
the data type \texttt{D} as example. The following is the type of the higher-order fold for \texttt{D}. 

\begin{verbatim}
hfoldD : {a b : Set} {p : Set -> Set -> Set} ->
         (dnil : {a b : Set} -> p a b) ->
         (dcons : {a b : Set} -> a -> b -> p (I a) b ->
                   p (p (I b) (I b)) (I a) -> p a b) ->
         (acons : {a b : Set} -> I b ->
                   p (I (I (p b a))) (p (p b a) (p a b)) -> p a b) ->
          D a b -> p a b
\end{verbatim}
To obtain this type, we first obtain \texttt{dnil}, \texttt{dcons} and \texttt{acons},
their types are the same as the types for the constructors \texttt{DNil}, \texttt{DCons} and \texttt{ACons}, except the type-constructor \texttt{D} is replaced by
the type variable \texttt{p} of kind \texttt{Set -> Set -> Set}.
We then use \texttt{bnil}, \texttt{bcons} and \texttt{acons} as the additional arguments for the function \texttt{D a b -> p a b}. Note that for any nested data type,
we can obtain the type of its higher-order fold this way.

%Now let us give the definition of \texttt{hfoldD} using \texttt{foldD}.
We now define the following type-level function \texttt{Hp} to replace the constructor \texttt{IsD} by
the binary type variable \texttt{p : Set -> Set -> Set}, while keeping the other
type-constructors unchanged. 

\begin{verbatim}
Hp : IndexD -> (Set -> Set -> Set) -> Set -> Set -> Set
Hp VarA p a b = a
Hp VarB p a b = b
Hp (IsD i j) p a b = p (Hp i p a b) (Hp j p a b)
Hp (IsI i) p a b = I (Hp i p a b)
\end{verbatim}

The higher-order fold \texttt{hfoldD} is defined by instantiating
 \texttt{foldD} with \texttt{\string\ i -> Hp i p a b}.

\begin{verbatim}
hfoldD {a} {b} {p} dnil dcons acons x =
  foldD {a} {b} {\ i -> Hp i p a b} (IsD VarA VarB) 
    (\ y -> y) (\ y -> y) dnil dcons acons Zero Succ x
\end{verbatim}
The higher-order fold \texttt{hfoldD} traverses over \texttt{D}, replacing
the constructors \texttt{DNil}, \texttt{DCons} and \texttt{ACons} by \texttt{dnil}, \texttt{dcons} and \texttt{acons}, while leaving the constructors \texttt{Zero} and \texttt{Succ} unchanged.  

\subsection{Discussion}
\label{most}
Looking back at the definition of \texttt{foldD}, it can be criticized for being too general. For example, although \texttt{foldD} is defined for folding \texttt{D},
we can also use \texttt{foldD} to define a summation for \texttt{I Nat}.

\begin{verbatim}
sumI : I Nat -> Nat
sumI l = foldD {Nat} {Nat} {\ y -> Nat} (IsI VarA) (\ y -> y) (\ y -> y)
          Z (\ x x1 x2 x3 -> Z) (\ x x1 -> Z) Z add l
\end{verbatim}
In this definition of \texttt{sumI}, we need to supply
additional arguments such as \texttt{\string\ x x1 x2 x3 -> Z} and \texttt{\string\ x x1 -> Z}
even though we know these arguments will not be used when evaluating \texttt{sumI}.

One way to understand why \texttt{foldD} may also work for \texttt{I} is that since the definition of \texttt{D} involves a nested use of \texttt{I}, so a fold for \texttt{I} is needed when folding \texttt{D}.
As a result, the function \texttt{foldD} may be used to operate on the values that only involves data type \texttt{I}. If one wants to program with the data type \texttt{I}, then it is more natural to program with the designated dependently typed fold for \texttt{I}.

We call the dependently typed folds obtained from the method in Section \ref{gen} \textit{the direct dependently typed folds}. The example of obtaining \texttt{foldD} shows that the direct
dependently typed folds always exist and we know how to systematically construct them.

In practice, we often have more intuition on how we intend to use
the nested data types. For example, consider the \texttt{Term} and \texttt{TermE} data type
in the previous sections, where we are particularly interested in types of the forms $\texttt{Term} \ (\texttt{Incr}^n a)$ and $\texttt{TermE} \ (\texttt{IncrTermE}^n a)$. So in these cases we use
 a natural number as the index to define folds for $\texttt{Term} \ (\texttt{Incr}^n a)$ and $\texttt{TermE} \ (\texttt{IncrTermE}^n a)$. Let us call the folds for $\texttt{Term} \ (\texttt{Incr}^n a)$ and $\texttt{TermE} \ (\texttt{IncrTermE}^n a)$ \textit{customized dependently typed folds}. 

Given a nested data type, its customized dependently typed folds can be different from its direct dependently typed fold. For example,
for the data types \texttt{Term} and \texttt{TermE}, the customized dependently typed folds \texttt{foldT} and \texttt{foldE} are not the same as the direct dependently typed folds.
On the other hand, for the data type \texttt{Bush}, the customized dependently typed fold \texttt{foldB} coincides with its direct dependently typed fold.

We say a dependently typed
fold is \textit{proper} if it can be specialized to the corresponding higher-order
fold. Because the higher-order fold is considered the \textit{defining property} of a nested
data type,
as the higher-order fold can be understood as the unique morphism from the initial nested data type object in the higher-order functor category \cite[\S 4]{bird1998nested}. We show that the direct dependently typed folds are proper in Section \ref{well-formed}. For the customized dependently typed folds, the properness has to be shown in a case by case basis (e.g. \texttt{hfoldT} and \texttt{hfoldE} for \texttt{Term} and \texttt{TermE}).

\section{Related Work}
\label{related}
%% (1) Works by Bird and companies.

%% (2) Works by Ghani.

%% (3) Works by Matthes.

\textbf{Generalized folds} This paper is inspired by the works of Bird, Paterson and Meertens. The higher-order folds such as \texttt{hfoldB} in Section \ref{intro}
were thought not expressive enough to define functions such as summation, which leads to the consideration of \textit{generalized folds} (\cite{bird1999bruijn}, \cite{bird1999generalised}). The generalized folds further generalize the existing higher-order folds with extra higher-order type variables and arguments. For example, the following is
a version of generalized fold for the \texttt{Bush} data type. We have to use the unsafe flag \texttt{--no-termination} to make Agda accept the following code.

{\small
\begin{verbatim}
gfoldB : {a : Set} -> {p q : Set -> Set} ->
        ({b : Set} -> p b) ->
        ({b : Set} -> q b -> p (p b) -> p b) ->
        ({b : Set} -> p b -> q (p b)) ->
        Bush (q a) -> p a
gfoldB base step k NilB = base
gfoldB {a} {p} {q} base step k (ConsB x xs) =
  step x (gfoldB {p a} {p} {q} base step k
           (hmapB (\ y -> k (gfoldB {a} {p} {q} base step k y)) xs))

hfoldB : {a : Set} -> {p : Set -> Set} ->
          ({b : Set} -> p b) -> ({b : Set} -> b -> p (p b) -> p b) -> Bush a -> p a
hfoldB {a} {p} base step = gfoldB {a} {p} {\ y -> y} base step (\ y -> y) 

sumB : Bush Nat -> Nat
sumB = gfoldB {Nat} {\ y -> Nat} {\ y -> Nat} Z add (\ x -> x) 
\end{verbatim}}
\noindent We can see the higher-order fold \texttt{hfoldB} is indeed an instance of the generalized fold \texttt{gfoldB}. Moreover, \texttt{gfoldB} can be
used to define functions such as \texttt{sumB}.

\textbf{Higher-order folds} Johann and Ghani show that the higher-order folds
such as \texttt{hfoldB} can be used to define functions such as \texttt{sumB} (\cite{johann2007initial}, \cite{johann2009principled}). The intuitive idea is instead of defining the summation function directly, one uses \texttt{hfoldB} to define the following auxiliary function \texttt{sumAux} first, and define \texttt{sumB'} based on \texttt{sumAux}.

{\small
\begin{verbatim}
sumAux : {a : Set} -> Bush a -> (a -> Nat) -> Nat
sumAux {a} = hfoldB {a} {\ a -> (a -> Nat) -> Nat} 
               (\ x -> Z) (\ x k f -> add (f x) (k (\ r -> r f)))  

sumB' : Bush Nat -> Nat
sumB' l = sumAux l (\ y -> y)
\end{verbatim}}
\noindent When defining \texttt{sumAux}, we instantiate the type variable \texttt{p} in \texttt{hfoldB} with \texttt{\string\ a -> (a -> Nat) -> Nat}. Johann and Ghani generalize the pattern of \texttt{sumAux} and show how it is related to an advanced concept called Kan extensions from category theory~\cite{johann2007initial}.

%% They also show that Church encodings can be derived from the higher-order folds for nested data type such as \texttt{Nested} in Section \ref{intro}, where the definition of such folds does not use map. It is unclear to us whether we can still derive Church encodings from the higher-order folds such as \texttt{foldB'}, as its definition uses map.   

\textbf{Comparison of dependently typed folds, generalized folds and higher-order folds}. We
already mentioned that dependently typed folds can be specialized to higher-order folds, and
that dependently typed folds does not requires map functions, and that map functions can be
defined by dependently typed folds, and that dependently typed folds are defined
using well-founded recursion, and that dependently typed folds correspond to the induction principles. %% Both generalized folds and higher-order folds require
%% general recursion and higher-order polymorphism, while
%% dependently typed folds requires well-founded recursion and dependent types.

%% Arguably, dependently typed folds enable a more direct style of programming with nested data types. 

 \textbf{Mendler-style iterators} Abel, Matthes and Uustalu propose to use a generalized version of Mendler-style iteration \cite{mendler1987inductive} to program with nested data types (\cite{abel2003generalized}, \cite{abel2005iteration}). The intuitive
idea is that one first defines nested data types as recursive types, i.e. the fix points of higher-order functors. For
example, the \texttt{Bush} data type is encoded as the following \texttt{Bush'} with the constructors \texttt{BNil} and \texttt{BCons}. 
%% \footnote{Note that the agda code here requires three unsafe flags \texttt{--no-termination, --no-positivity, --type-in-type}.}.

{\small
\begin{verbatim}
data Mu (F : (Set -> Set) -> (Set -> Set)) (a : Set) : Set where
  In : F (Mu F) a -> Mu F a
BushF : (Set -> Set) -> (Set -> Set)
BushF B a = Unit + a * (B (B a))
Bush' : Set -> Set
Bush' = Mu BushF
BNil : {a : Set} -> Bush' a
BNil = In (Inl unit)
BCons : {a : Set} -> a -> Bush' (Bush' a) -> Bush' a
BCons x xs = In (Inr (Times x xs))
\end{verbatim}}
\noindent Note that the data type \texttt{Mu} is not strictly-positive in Agda, we have to
use the unsafe flag \texttt{--no-positivity} to make Agda accept the above code.

One then defines the following generalized Mendler-style iterator \texttt{gIt}. The iterator \texttt{gIt} uses a generalized impredicative type abstraction \texttt{mon}. 

{\small
\begin{verbatim}
mon : (Set -> Set) -> (Set -> Set) -> (Set -> Set) -> Set
mon F H G = {a b : Set} -> (a -> H b) -> F a -> G b

gIt : {F : (Set -> Set) -> (Set -> Set)} {H G : Set -> Set} ->
      ({X : Set -> Set} -> mon X H G -> mon (F X) H G) ->
      mon (Mu F) H G
gIt s f (In t) = s (gIt s) f t
\end{verbatim}}
\noindent We have to use the unsafe flag \texttt{--type-in-type} to make Agda accept the
definition of \texttt{mon}. Moreover, another unsafe flag \texttt{--no-termination} is needed
because the termination of \texttt{gIt} is not obvious for Agda. 

The \texttt{sumAux'} function and the map function \texttt{mapBush} can be defined from \texttt{gIt}.
%Note that the map function for \texttt{Bush} can also be defined from \texttt{gIt}.

{\small
\begin{verbatim}
sumAux' : {a : Set} -> (a -> Nat) -> Bush' a -> Nat
sumAux' {a} = gIt {BushF} {\ x -> Nat} {\ x -> Nat}
               (\ {X} r {a} {b} f t ->
                   match t (\ x -> Z)
                           (\ p -> add (f (p1 p))
                                   (r {X a} {b} (r {a} {b} f) (p2 p)))) {a} {a} 
mapBush : {X Y : Set} -> (X -> Y) -> (Bush' X -> Bush' Y)
mapBush = mfold bushF In         
\end{verbatim}}

Although in Agda, the \texttt{Mu} data type is not strictly positive, the definition of
\texttt{mon} requires impredicative polymorphism and the definition of
\texttt{gIt} is not obviously terminating, Abel et. al. \cite{abel2003generalized} show that the recursive types together with the generalized Mendler-style iterators can be encoded in Girard's $\textbf{F}_\omega$ \cite{girard1972interpretation} using a syntactic version of Kan extensions, hence the programs defined by the generalized Mendler-style iterators are still terminating.

%\cite{wadler1990recursive}
\textbf{Induction principles for Mendler-style iterators} Matthes proposes to
use a system called \textit{LNMIt} (logic for natural Mendler-style iteration of rank 2) \cite[Fig. 1.]{matthes2009induction} to reason about programs defined by Mendler-style iterators. The \textit{LNMIt} consists of the usual constants such as \texttt{In} and \texttt{gIt}, it also contains the special constants for map and induction. Although the map constant and the iterator \texttt{gIt} come with reduction rules, there is no such rule for the induction constant. The type of the induction constant contains the map constant, and the induction constant provides
a mean to show a property holds for the data types defined from \texttt{Mu}. It is unclear to us how the induction constant in \textit{LNMIt} is related to the iterator \texttt{gIt}. Matthes \cite[\S 5]{matthes2009induction} proves in Coq that \textit{LNMIt} can be
defined within the Calculus of Inductive Construction \cite{coquand1990inductively} with the additional axioms of impredicative \texttt{Set} and proof irrelevance. 

\textbf{Comparison of dependently typed folds, generalized Mendler-style iterators and LNMIt}.
Both dependently typed folds and generalized Mendler-style iterators enable total
programming with iterators, and allow maps functions to be defined from the iterators.
The approach of generalized Mendler-style iterators requires working with recursive types instead of the usual inductive definitions, it has the advantage of not imposing any
positivity constraint for the data types. The dependently typed folds approach requires working with the inductive definitions of data types, it is limited to a subclass of strictly positive data types. 
%% The type system feature that generalized Mendler-style iterators need is quite minimal, namely, Girard's System $\textbf{F}_{\omega}$ is enough to encode the generalized Mendler-style iterators.
%% Using the index representations and Church encodings, dependently typed folds can be carried out in
%% Calculus of Construction \cite{coquand1988calculus}, which is an extension of $\textbf{F}_{\omega}$.
As for
the verification of programs involving nested data types, the justification of System \textit{LNMIt} requires impredicative \texttt{Set} and proof irrelevance, while the induction principles obtained
from dependently typed folds work directly in total Agda and does not requires the axioms of impredicative \texttt{Set} and proof irrelevance.

\textbf{Other related work from type theory}
The main technique we use to realize dependently typed folds is called \textit{large elimination} in the dependent types literature (\cite{alti:phd93}, \cite{werner1992normalization}), i.e.
computing types by pattern-matching on values. 
Werner shows that the inductive reasoning in the Calculus of Inductive Construction is consistent \cite{werner:tel-00196524}. Modern
dependently typed languages such as Agda, Coq allow user-defined regular/nested data types and well-founded recursive function definitions \cite{coquand1992pattern}, this enables us to define the dependently typed folds, their corresponding induction principles and various type-level functions. 

%% in  one 

\section{Conclusion and future work}
\label{final}
We show how to define dependently typed folds for nested data types and how to specialize them to the corresponding higher-order folds. Dependently typed folds can be used to define maps, and
 other terminating functions. They give rise to the induction principles, similar to the folds for regular data types. We show 
how to use induction principles to reason about the programs involving nested data types.
We also discuss how dependently typed folds give rise to the indexed representations and how to obtain the Church encodings of the indexed representations.

For future work, we would like to formalize a general procedure to obtain the direct
dependently typed folds. We would also like to consider the meta-theoretic properties of
dependently typed folds. For example, it would be nice to be able to prove a meta-theorem
stating that any higher-order fold obtained from the direct dependently typed fold will
have the same computational behavior as the higher-order fold obtained from general recursion.

\bibliography{dep}

%% Appendix
\appendix

\end{document}